\documentclass[aps,pra,reprint,superscriptaddress]{revtex4-1}

\usepackage{amsmath}
\usepackage{graphicx}
\usepackage{amsfonts}
\usepackage{color}
\usepackage{bm}
\usepackage{lipsum}
\usepackage{comment}

\usepackage{amsmath}
\usepackage{amssymb}
\usepackage{graphicx} 
\usepackage{braket}
\usepackage{mathtools}
\usepackage{times}
\usepackage{enumerate}

\usepackage[colorlinks=true,linkcolor=blue,urlcolor=blue,citecolor=blue,pdfusetitle]{hyperref}

\graphicspath{{images/}}

\newcommand{\Tr}{\mathrm{Tr}}

\newcommand*{\defeq}{\mathrel{\vcenter{\baselineskip0.5ex\lineskiplimit0pt\hbox{\scriptsize.}\hbox{\scriptsize.}}}=}

\begin{document}

\title{Entropy production in the mesoscopic-leads formulation of quantum thermodynamics}

\date{\today}

\author{Artur M. Lacerda}
\email{machadoa@tcd.ie}
\affiliation{School of Physics, Trinity College Dublin, College Green, Dublin 2, D02K8N4, Ireland}

\author{Michael J. Kewming}
\affiliation{School of Physics, Trinity College Dublin, College Green, Dublin 2, D02K8N4, Ireland}

\author{Marlon Brenes}
\affiliation{Department of Physics and Centre for Quantum Information and Quantum Control, University of Toronto, 60 Saint George St., Toronto, Ontario, M5S 1A7, Canada}
\affiliation{Centro de Investigación en Ciencia e Ingeniería de Materiales (CICIMA), Universidad de Costa Rica, San José, Costa Rica}
\affiliation{Escuela de Física, Universidad de Costa Rica, San José, Costa Rica}

\author{Conor Jackson}
\affiliation{H. H. Wills Physics Laboratory, University of Bristol, Bristol, BS8 1TL, United Kingdom}

\author{Stephen R. Clark}
\affiliation{H. H. Wills Physics Laboratory, University of Bristol, Bristol, BS8 1TL, United Kingdom}

\author{Mark T. Mitchison}
\email{mark.mitchison@tcd.ie}
\affiliation{School of Physics, Trinity College Dublin, College Green, Dublin 2, D02K8N4, Ireland}
\affiliation{Trinity Quantum Alliance, Unit 16, Trinity Technology and Enterprise Centre, Pearse Street, Dublin 2, D02YN67, Ireland}

\author{John Goold}
\email{gooldj@tcd.ie}
\affiliation{School of Physics, Trinity College Dublin, College Green, Dublin 2, D02K8N4, Ireland}
\affiliation{Trinity Quantum Alliance, Unit 16, Trinity Technology and Enterprise Centre, Pearse Street, Dublin 2, D02YN67, Ireland}

\begin{abstract}
Understanding the entropy production of systems strongly coupled to thermal baths is a core problem of both quantum thermodynamics and mesoscopic physics. While there exist many techniques to accurately study entropy production in such systems, they typically require a microscopic description of the baths, which can become numerically intractable to study for large systems. Alternatively an open-systems approach can be employed with all the nuances associated with various levels of approximation. Recently, the mesoscopic leads approach has emerged as a powerful method for studying such quantum systems strongly coupled to multiple thermal baths. In this method, a set of discretised lead modes, each locally damped, provide a Markovian embedding. Here we show that this method proves extremely useful to describe entropy production of a strongly coupled open quantum system. We show numerically, for both non-interacting and interacting setups, that a system coupled to a single bath exhibits a thermal fixed point at the level of the embedding. This allows us to use various results from the thermodynamics of quantum dynamical semi-groups to infer the non-equilibrium thermodynamics of the strongly coupled, non-Markovian central systems. In particular, we show that the entropy production in the transient regime recovers the well established microscopic definitions of entropy production with a correction that can be computed explicitly for both the single- and multiple-lead cases.

\end{abstract}

\maketitle

\section{Introduction}

Many non-equilibrium systems, ranging from our planet and biological systems to the quantum dynamics of electrons in nano-structures, are driven by thermodynamic affinities and dissipate energy to the environment. 
This is the process of entropy production, which is the characteristic thermodynamic signature of an irreversible transformation \cite{Landi_2021, Strasberg_tutorial_2021} and also the focal point of non-equilibrium thermodynamics \cite{Lebowitz_statistical_1999}. 
A physical scenario of particular interest is where a central system is driven out of equilibrium by thermodynamic reservoirs and/or external drives that generate currents of particles and heat which flow and fluctuate in time, producing entropy.
In the nanoscale domain, where boundary effects are dominant, the central system may be strongly coupled to these reservoirs and heavily influenced by their non-trivial spectral properties. 
In such a configuration, there are only a few approaches for studying the entropy production---in both transient and stationary states---other than constructing a full microscopic description of the joint system plus the affinities~\cite{Landi_2021,Talkner2020}.

In these configurations, one can show that the irreversible entropy production equals the relative entropy between the initial and final reduced states of the reservoirs, plus the mutual information between the reservoirs and the central system \cite{Esposito_2010, Reeb_2014, Strasberg_quantum_2017, Ptaszynksi_entropy_2019}.
The connection between the quantum relative entropy and the entropy production has been well established in many other works \cite{Chen_thermodynamic_2017, Wen_production_2017, Engelhardt_2018, Manzano_2018, Santos_2019, Bera_2019}, but its central role in strong coupling was formalised most succinctly for fermionic leads in Ref.~\cite{Ptaszynksi_entropy_2019}, where intraenvironment couplings were clearly accounted for. This approach typically requires a full treatment of the unitary evolution between system and leads but we show here that it can be obtained also from a master equation in Lindblad form where the system and a damped discretisation of the leads form a Markovian embedding~\cite{Woods2014}.  
This allows us to apply the seminal result by Spohn who recognised that entropy production can be defined in terms of the quantum relative entropy for an arbitrary quantum dynamical semigroup with a stationary state \cite{Spohn_1978}. The power of the Markovian embedding is that it allows us to access the entropy production of a non-Markovian system in the strong coupling regime by extending the system through this embedding. We stress that this approach allows us to avoid the known thermodyamic pitfalls that are associated with Lindblad master equations that are often used in quantum thermodynamics~\cite{Walls1970, Carmichael1973, Wichterich_2007, Rivas_2010, Levy2014, Barra2015, Trushechkin_2016, Gonzalez_2017, Hofer_2017, Naseem2018, Chiara2018, Mitchison_2018, Purkayastha2022}.

Here, we employ the mesoscopic leads approach, a powerful technique which can be used to obtain the dynamics of the system in strong-coupling or non-Markovian regimes \cite{Imamoglu_1994,Garraway_1997a,Garraway_1997b,Sanchez_2006,Subotnik_2009,Dzhioev_2011, Ajisaka_2012,Ajisaka_2013,Arrigoni_2013,Dorda_2014,Chen_2014,Zelovich_2014,
Dorda_2015,Hod_2016,Gruss_2016,Schwarz_2016,Dorda_2017,Elenewski_2017,
Gruss_2017,Zelovich_2017,Tamascelli_2018,Lemmer_2018,Oz_2019,Chen_Galperin_2019,
Zwolak_2020,Wojtowicz_2020,Chiang_2020,Brenes2020,Lotem_2020,Fugger_2020,
Wojtowicz_2021,Elenewski_2021}. 
The power of this method stems from the approximation made on the macroscopic baths.
In this approach, they are systematically approximated by a finite number of damped modes. 
Initially introduced for bosonic baths \cite{Imamoglu_1994,Garraway_1997a,Garraway_1997b}, this approach has subsequently been used extensively to study quantum transport in fermionic set-ups also \cite{Sanchez_2006,Subotnik_2009,Dzhioev_2011,
Ajisaka_2012,Ajisaka_2013,Arrigoni_2013,Dorda_2014,Chen_2014,Zelovich_2014,
Dorda_2015,Hod_2016,Gruss_2016,Schwarz_2016,Dorda_2017,Elenewski_2017,
Gruss_2017,Zelovich_2017,Chen_Galperin_2019,Chen_Galperin_2019_2,Oz_2019,
Zwolak_2020,Wojtowicz_2020,Chiang_2020,Brenes2020,Lotem_2020,Fugger_2020,
Wojtowicz_2021,Elenewski_2021}. 
When combined with tensor network techniques, it has been recently shown that it is possible to completely obtain energy and particle currents in non-equilibrium steady states (NESS) of interacting quantum many-body systems, and thereby the thermodynamics at NESS \cite{Brenes2020}. 
Other works have also used this approach to describe impurity models at and beyond Kondo regimes \cite{Arrigoni_2013,Dorda_2014,Dorda_2015,Dorda_2017,Lotem_2020,Fugger_2020}. Further powerful applications of the mesoscopic leads approach are the study of time-dependent system Hamiltonians \cite{Chen_2014,Oz_2019, Lacerda_2022} and full counting statistics of the particle current \cite{Brenes_2022}.

Our objective here is show that the mesoscopic leads approach can accurately compute the entropy production of a strongly coupled system without having to employ a full unitary dynamical description at the level of system and baths. We begin by first introducing the mesoscopic leads approach and the Lyapunov equation for Gaussian systems (Sec.~\ref{sec:mesoleads}) and show how both the relative entropy and the fidelity can be expressed between two Gaussian states in terms of their covariance matrices. We then discuss our numerical scheme for extending to non-quadratic systems. 
We present strong numerical evidence that the fixed point of the Markovian embedding is a thermal distribution (Sec.~\ref{sec:thermalisation}) in both Gaussian and non-Gaussian cases. Our results are fully consistent with other recent studies of thermalisation in open quantum systems~\cite{Reichental2018, Zanoci2023}.
Building on this, we then show that the mesoscopic leads approach agrees with Spohn's result \cite{Spohn_1978} for the transient entropy production rate of a system coupled to single reservoir (Sec.~\ref{sec:spohn}) at the level of the embedding.
We finally derive analytically an expression for the difference in cumulative (time-integrated) internal and external entropy productions in the mesoscopic leads approach (Sec.~\ref{sec:difference_between_internal_and_external}). This derivation is performed in the single- and multiple-bath cases, clarifying when the two definitions of entropy production agree. The approach can be easily generalised to arbitrary bath spectral densities and time-dependent system Hamiltonians. 

\section{The Mesoscopic Leads Approach}
\label{sec:mesoleads}

\begin{figure}
    \centering
    \includegraphics{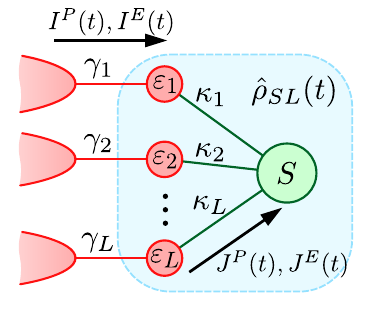}
    \caption{A system $S$ coupled to a bath can be approximated by $L$ modes which are each independently coupled to independent infinite baths. Each lead mode, with on-site energy $\varepsilon_{k}$, is coupled to the central system $S$ with coupling strength $\kappa_{k}$. The lead is damped by infinite bath with decay rate $\gamma_{k}$. We can define the internal currents $J^{P}(t)$ and $J^{E}(t)$, and the external currents $I^{P}(t)$ and $I^{E}(t)$.}
    \label{fig:mesoleads}
\end{figure}

Through the mesoscopic leads approach, a thermal reservoir is discretised into a collection of $L$ fermionic modes with energies $\varepsilon_k$ for the $k$-th mode, each of which is coupled to \emph{residual} reservoirs. Each residual reservoir brings each fermionic mode to its energy-dependent thermal equilibrium state and, crucially, we shall assume that each residual reservoir is Markovian. The configuration couples to a quantum system $S$ generically and, in such a way, the \emph{extended system} composed of $S$ and the lead modes evolves under Markovian dynamics (see Fig.~\ref{fig:mesoleads}). We denote the system Hamiltonian as $\hat{H}_S$, written in terms of canonical fermionic operators $\{ \hat{c}_j \}$ for the $j$-th fermionic site.
In this section we introduce the most important aspects of the formulation. We refer interested readers to Refs.~\cite{Brenes2020, Brenes_2022} for explicit microscopic derivations. Units where $\hbar = 1$ and $k_B=1$ are used throughout.

\subsection{The extended system}
For the sake of simplicity, we shall introduce this formulation for the case of a single reservoir coupled to the system. The same formalism can be extended trivially to multi-reservoir configurations.  

The Hamiltonian of the extended system is given by 
\begin{align}
\label{eq:hamiltonian_maps_to}
    \hat{H} = \hat{H}_{S} + \hat{H}_{L} + \hat{H}_{SL},
\end{align}
where $\hat{H}_{L} = \sum_{k=1}^{L} \varepsilon_k \hat{a}^{\dagger}_{k} \hat{a}_{k}$ is the \emph{lead} Hamiltonian and $\hat{H}_{SL}$ is the interaction Hamiltonian between the system and lead modes, given by
\begin{align}
    \hat{H}_{SL} = \sum_{k=1}^{L} \kappa_{k} \big( \hat{c}_p^\dagger \hat{a}_{k} + \hat{a}_{k}^\dagger \hat{c}_p\big),
\end{align}
where we have assumed that the lead couples locally to the $p$-th site of the system, as in Fig.~\ref{fig:mesoleads}. In our formulation, there are no terms that couple system and residual-reservoir degrees of freedom, however, they do interact though the lead modes.

Each of the lead modes couples to a residual reservoir with lead-reservoir coupling strength $\gamma_k$. Crucially, the lead-reservoir couplings $\gamma_k$ are assumed to be weak, such that one can justify and microscopically derive a Markovian master equation of the GKLS type~\cite{Brenes2020, Brenes_2022} that dictates the dynamics of the system plus the lead modes. Denoting $\hat{\rho}_{SL}(t) $ as the time-dependent state of the system plus lead modes, one finds (in the Schr\"odinger picture)
\begin{align}
\label{eq:master_equation}
    \frac{{\rm d}\hat{\rho}_{SL}(t)}{{\rm d}t} &= \mathcal{L}\{\hat{\rho}_{SL} (t)\} \nonumber \\
    &\defeq {\rm{i}}[\hat{\rho}_{SL}(t),\hat{H}_{S} + \hat{H}_{L} + \hat{H}_{SL}] + \mathcal{D}\{\hat{\rho}_{SL}(t)\}.
\end{align}
The dissipators are given by 
\begin{align}
\label{eq:dissipator}
\mathcal{D}\{\hat{\rho}\} = &\sum_{k=1}^{L} \gamma_{k}(1 - f_{k}) \left[\hat{a}_{k} \hat{\rho} \hat{a}^{\dagger}_{k} - \tfrac{1}{2}\{ \hat{a}^{\dagger}_{k} \hat{a}_{k}, \hat{\rho} \} \right] \nonumber \\
& + \sum_{k=1}^{L} \gamma_{k} f_{k} \left[\hat{a}^{\dagger}_{k} \hat{\rho} \hat{a}_{k} - \tfrac{1}{2}\{ \hat{a}_{k} \hat{a}^{\dagger}_{k}, \hat{\rho} \} \right].
\end{align}
We have defined $f_{k} \defeq f(\varepsilon_{k}) = (e^{\beta(\varepsilon_{k}-\mu)} + 1)^{-1}$ as the energy-dependent Fermi-Dirac distribution for the lead modes, which is parameterised by the inverse temperature $\beta$ and chemical potential $\mu$.

The couplings $\kappa_{k}$ between each lead mode and the $p$-th system site dictate the effective spectral function of the bath acting on the central system. In principle, one can construct any continuous spectral function using this formulation~\cite{Brenes2020, Brenes_2022}. However, we shall focus on a flat (wide-band) spectral function
\begin{align}
\label{eq:wideband}
    \mathcal{J}(\omega) = \begin{cases} \Gamma,\; \forall\, \omega \in [-W, W] \\ 0,\; \textrm{otherwise}, \end{cases}
\end{align}
where $W$ is some cut-off energy scale and $\Gamma$ is the effective or average coupling to the reservoir. It can be shown~\cite{Dzhioev_2011} that for this specific choice of spectral function, $\Gamma = 2\pi \kappa_{k}^2/e_{k}$ in the wide-band limit above, where $e_{k} = \varepsilon_{(k+1)} - \varepsilon_{k}$. In this sense, we discretise the reservoir into $L$ energy modes $\varepsilon_k$ between $-W$ and $W$, such that $e_{k} = 2W / L$. The external coupling between each energy mode in the leads and their own residual environment $\gamma_{k}$ is in principle arbitrary, as long as it remains a perturbative energy scale in the problem so that the Lindblad equation [Eq.~\eqref{eq:master_equation}] holds. Here we take $\gamma_{k} = e_{k} = 2W / L$, thus obtaining a controlled approximation as $L \to \infty$~\cite{Zwolak1:2021,Zwolak2:2021}. It is important to remark that even though the $\gamma_{k}$ are small parameters in the energy scales of the configuration, the internal coupling $\Gamma$ is not necessarily so and, in such a way, the strong system-reservoir coupling regime may be addressed. Finally, we note that Eq.~\eqref{eq:master_equation} can be trivially extended to multi-reservoir configurations by adding the corresponding dissipators. 

\subsection{Non-interacting fermionic systems: fidelity and relative entropy}
\label{subsec:lyapunov_eq}

Thus far, we have made no assumptions about the system Hamiltonian $\hat{H}_S$. We shall now consider a time-independent central system comprising $N$ fermionic sites with nearest-neighbour tunnelling, such that
\begin{align}
\label{eq:h_s}
   \hat{H}_{S} = \sum_{j=1}^{N}\epsilon_j \hat{c}^{\dagger}_j \hat{c}_j - g\sum_{j=1}^{N-1} \left( \hat{c}^{\dagger}_j \hat{c}_{j+1} + \hat{c}^{\dagger}_{j+1} \hat{c}_j \right)\,,
\end{align}
where $\epsilon_{k}$ is the on-site energy of the system and $g$ is the coupling between each adjacent site. 
We will also assume the lead is coupled to the system in the first site, i.e., $p = 1$.

For this specific choice of $\hat{H}_{S}$, the quadratic form of $\hat{H}$ indicates that Eq.~\eqref{eq:master_equation} can be solved via a Lyapunov-type equation~\cite{Brenes_2022}. Note, however, that the bases of operators are mixed: $\hat{H}_{L}$ is defined in terms of canonical operators in the energy basis, while $\hat{H}_{S}$ is defined through operators in the spatial configuration basis. In this regard, the total Hamiltonian of a system composed of $K$ leads, each with $L$ modes, and a tight-binding Hamiltonian of $N$ sites, can be written compactly in the form 
\begin{align}
\hat{H} = \sum_{i,j}^{KL+N} [\mathbf{H}]_{ij} \hat{d}_i^\dagger \hat{d}_j,
\end{align}
where $i, j = 1,2,\cdots,KL+N$, and $\mathbf{H}$ is a hermitian matrix. We use $\hat{d}$ and $\hat{d}^{\dagger}$ operators to label both system and lead operators, while keeping in mind that the basis is mixed between configurational degrees of freedom for system operators and energy degrees of freedom for lead operators.

We can now define a covariance matrix $\mathbf{C}$ with entries 
\begin{equation}
\label{eq:covariance}
    [\mathbf{C}]_{ij} = C_{ij} =  \textrm{Tr}[\hat{\rho}_{SL} \hat{d}_j^\dagger \hat{d}_i].
\end{equation}
The covariance matrix obeys the equation of motion
\begin{align}
\label{eq:lyapunov}
    \frac{{\rm d}\mathbf{C}(t)}{{\rm d}t} = -\left(\mathbf{W}\mathbf{C} + \mathbf{C}\mathbf{W}^{\dagger} \right) + \mathbf{F}
\end{align}
where 
\begin{equation}
    \mathbf{W} = {\rm{i}} \mathbf{H} + \frac{\bm{\gamma}}{2}, 
    \qquad 
    [\mathbf{F}]_{kk} = F_k = \gamma_k f_k,
\end{equation}
with $\bm{\gamma}$ being a diagonal matrix with entries $[\bm{\gamma}]_{kk} = \gamma_k$.
Eq.~\eqref{eq:lyapunov} gives a closed-form expression for the dynamics of the covariance matrix for Gaussian systems~\cite{Brenes_2022}. 
For the thermalisation configuration, we are only interested in the long-time solution to the covariance matrix, i.e., the steady state in which 
\begin{align}
    \frac{{\rm d}\mathbf{C}(t)}{{\rm d}t} = \mathbf{0}.
\end{align}
Then, in the steady state, the covariance matrix is the solution to the equation
\begin{align}
\label{eq:Css}
    \left(\mathbf{W}\mathbf{C}_{ss} + \mathbf{C}_{ss}\mathbf{W}^{\dagger} \right) = \mathbf{F}\,.
\end{align}

We can use the covariance matrix to compute other quantities that will be of interest in the proceeding sections. First, the relative entropy between two Gaussian states $D(\hat{\rho}_{1}||\hat{\rho}_{2}) = - S(\hat{\rho}_{1}) - \Tr(\hat{\rho}_{1} \log \hat{\rho}_{2})$ where $S(\hat{\rho}) = - \Tr(\hat{\rho}\log \hat{\rho})$ is the von Neumann entropy, and second, the quantum fidelity $F(\hat{\rho}_{1}||\hat{\rho}_{2}) = {\rm Tr}\left( \sqrt{\hat{\rho}_{1}\hat{\rho}_{2}}\right)$.

We will rely on fact that the density matrix of a Gaussian system is uniquely defined as \cite{Peschel_2003, Bravyi2005, Dhar2012}
\begin{align}
\label{eq:Gaussian_rho}
    \hat{\rho} = \frac{e^{-\mathbf{d}^{\dagger} \mathbf{M}\mathbf{d}}}{Z}\,,
\end{align}
where $\mathbf{d}$ is a vector comprising all the system-lead operators $\{ \hat{d}_i \}$.
The matrix $\mathbf{M}$ and the partition function $Z$ are, respectively, given by
\begin{align}
    \mathbf{M} = \log\left({\frac{\mathbf{1} - \mathbf{C}}{\mathbf{C}}}\right),
    \qquad
    Z = \frac{1}{\textrm{det}[\mathbf{1} - \mathbf{C}]}\,.
\end{align}
This is a particularly useful representation, as discussed in Appendix~\ref{app:Fidelity and Entropy calculation}. Eq.~\eqref{eq:Gaussian_rho} can be used to derive the von Neumann entropy
\begin{align}
    S(\hat{\rho}) = \log(Z) + \textrm{Tr}[\mathbf{M}\mathbf{C}]\,.
\end{align}
Furthermore, the relative entropy between two Gaussian states $\hat{\rho}_{1}$ and $\hat{\rho}_{2}$ is given by 
\begin{align}
\label{eq:Gaussian_relative_entropy}
    D(\hat{\rho}_{1}|| \hat{\rho}_{2}) &= -S(\hat{\rho}_{1}) + \log(Z_{2}) + \textrm{Tr}[\mathbf{M}_{2} \mathbf{C}_{1}]\,,\\
    &=\log \frac{Z_{2}}{Z_{1}} + {\rm Tr}\left([\mathbf{M}_{2} - \mathbf{M}_{1}]\mathbf{C}_{1}\right)\,,
\end{align}
where $\mathbf{M}_j$, $\mathbf{C}_j$, and $Z_j$ pertain to the Gaussian state $\hat{\rho}_j$.
The quantum fidelity between $\hat{\rho}_{1}$ and $\hat{\rho}_{2}$ can be computed as 
\begin{equation}
\label{eq:fidelity}
    F(\hat{\rho}_{1}||\hat{\rho}_{2}) = {\rm Tr}\left( \sqrt{\hat{\rho}_{1}\hat{\rho}_{2}}\right) =  \frac{\det(\mathbf{1} + e^{-\frac{1}{2}\mathbf{M}_1}e^{-\frac{1}{2}\mathbf{M}_2})}{\sqrt{Z_1 Z_2}}\,,
\end{equation}
which can be used a measure for the distance between two states. We note that a similar result was first derived in Ref.~\cite{Banchi_quantum_2014}.

\begin{figure*}
    \centering
    \includegraphics{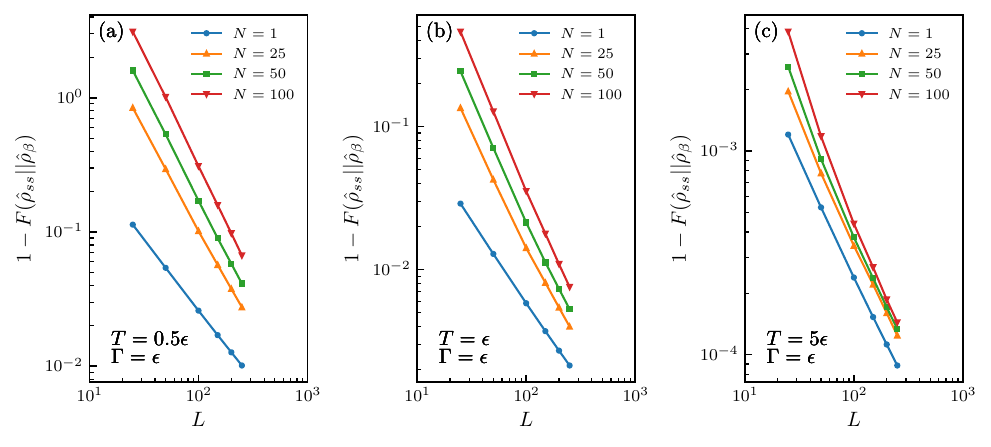}
    \caption{The infidelity, $1-F(\hat{\rho}_{ss}||\hat{\rho}_\beta)$ between the canonical state $\hat{\rho}_{\beta}$ of the enlarged system plus leads configuration [see Eq.\eqref{eq:thermal}] and the steady state $\hat{\rho}_{ss}$. The system comprises a nearest-neighbour fermionic chain of $N$ sites with the single edge mode coupled to $L$ lead modes. We chose a coupling coefficient $g=\epsilon = \epsilon_j$ [see Eq.~\eqref{eq:h_s}]. The infidelity is computed using Eq.~\eqref{eq:Gaussian_relative_entropy}. We selected $T = 0.5\epsilon, \epsilon, 5\epsilon$ [panels (a), (b) and (c)].}
    \label{fig:infidelity}
\end{figure*}

\subsection{Interacting fermionic systems: tensor networks and the variability}
\label{sec:interacting}

We will also address interacting systems, which elude solution by the preceding methods. 
As an example, we will examine systems governed by the nearest-neighbour Hamiltonian, Eq.~\eqref{eq:h_s}, but with an additional density-density interaction between adjacent sites:
\begin{align}
\label{eq:heisenberg_interacting}
       \hat{H}_{S} = \sum_{j=1}^{N}\epsilon_j \hat{c}^{\dagger}_j \hat{c}_j - \sum_{j=1}^{N-1} g (\hat{c}^{\dagger}_j \hat{c}_{j+1}+{\rm{H.c}}) + \sum_{j=1}^{N-1} U\hat{n}_{i} \hat{n}_{i+1}.
\end{align}
The additional term proportional to $U$ introduces a quartic interaction that cannot be modelled using the covariance matrix alone. 

Instead, to calculate the thermal and steady states, we employ a matrix product state (MPS) based approach based on the superfermion description of the entire Markovian embedding~\cite{Brenes2020, Dzhioev_2011}. The superfermion approach transforms the density matrix of $N + L$ sites into a pure state on $2N + 2L$ sites, pairing each physical site with an ancilla. This method offers the advantage of incorporating fermionic correlations without introducing cumbersome long-range Jordan-Wigner strings~\cite{Brenes_2022}. 

From tensor network calculations, it is very complicated to compute quantities which depend on high powers of the density matrix, such as the fidelity. Nevertheless, linear or quadratic functions of the density matrix, such as inner products and expectation values, can be readily computed. Therefore, we rely on the Hilbert-Schmidt norm of the action of the Lindbladian on a given state, denoted as the \emph{variability}
\begin{equation}
    v(\hat{\rho}) = \sqrt{\Tr{\mathcal{L}\{\hat\rho\}^{\dagger}\mathcal{L}\{\hat\rho\}}}\,.
\end{equation}
This measure is motivated by the fact that $\mathcal{L}{\rho_{ss}} = 0$, where $\hat{\rho}_{ss}$ is the steady state (or fixed point).  Therefore, if $v(\hat{\rho}_\beta) \approx 0$, then $\hat{\rho}_\beta$ is an approximate fixed point of the dissipative dynamics. In particular, we will show this to hold when $\hat{\rho}_\beta$ is a thermal state of the entire extended system at the externally imposed temperature $T = 1/\beta$ of the lead (see Sec.~\ref{sec:interacting_thermalisation}). This approach has the significant advantage that it only requires knowledge of the thermal state and does not require a separate calculation of the steady state.  Note that we assume throughout that the steady state is unique, which is justified by the lack of any known strong symmetries~\cite{Buca2012} and the fact that we consider ergodic models in the interacting case.

A key technical aspect should be highlighted about the MPS calculation described above. It is usually the case that the non-quadratic term in Eq.~\eqref{eq:heisenberg_interacting} leads to the build-up of strong correlations between the system and the bath, which leads to diverging bond dimensions in the MPS ansatz employed for both time evolution and the computation of steady-states for interacting systems. However, the effect of the Markovian embedding is such that these correlations are curtailed by the action of the Lindblad dissipators on the extended system. This allows us to study strong interactions and system-bath couplings using MPS approaches with tractable bond dimensions. We refer the reader to Refs.~\cite{Brenes2020,Cirio:2023Pseudo} for further details.

\section{Thermalisation and Entropy production in Mesoleads}

\subsection{Thermalisation of the extended system steady state}
\label{sec:thermalisation}
\subsubsection{Non-interacting systems}

As a starting point, we would like to determine whether the steady state of the extended system $\rho_{SL}(t\rightarrow \infty)$ is a thermal state when it is  coupled to a single bath. We focus first on the non-interacting configuration described by Eq.~\eqref{eq:h_s}, assuming that the lead
is coupled to the system via the first site. This involves evaluating the solution of Eq.~\eqref{eq:Css}, which yields $\mathbf{C}_{ss}$.

To this end, we can answer this question by comparing the steady-state covariance matrix with that of a thermal state $\hat{\rho}_{\beta}$, which is given by the density matrix 
\begin{align}
\label{eq:thermal}
    \hat{\rho}_{\beta} = \frac{e^{-\beta (\hat{H} - \mu \hat{N})}}{Z},
\end{align}
where $\hat{H} = \hat{H}_{S} + \hat{H}_{SL} + \hat{H}_{L}$ and $\hat{N} = \sum_{k=1}^{N} \hat{c}^\dagger_{k} \hat{c}_{k}+  \sum_{k=1}^{L} \hat{a}^\dagger_{k} \hat{a}_{k} $ are the Hamiltonian and total number operator of the extended system, respectively. For the sake of simplicity and without loss of generality, we shall consider $\mu = 0$.

In Fig.~\ref{fig:infidelity} we show the infidelity, $1-F(\hat{\rho}_{ss} || \hat{\rho}_{\beta})$, as a function of the number of lead modes, $L$. In Fig.~\ref{fig:infidelity} we display this calculation for different values of temperature and system sizes $N$ [see Eq.\eqref{eq:h_s}]. It can be observed that, as the number of lead modes $L$ increases, the infidelity monotonically approaches zero. This is an indication that the extended system is approaching a thermal state. We note that a larger number of lead modes $L$ are required to achieve lower infidelity values as the number of system modes $N$ increases. 

\subsubsection{Interacting systems}\label{sec:interacting_thermalisation}

To show that the fixed point is thermal in the interacting case, we employ the MPS-superfermion approach outlined in Sec.~\ref{sec:interacting}. To generate a thermal state $\hat{\rho}_\beta$, we use imaginary time evolution following the time-dependent variational principle (TDVP) with a time step of $\Delta \beta = 0.001$, an MPS cutoff of $10^{-6}$, a maximum bond dimension of $80$, and a minimum bond dimension of $15$.
Thermal states for a given Hamiltonian are computed in a single sweep from infinite temperature, progressing from the highest temperature state to the lowest.

\begin{figure}
    \centering
    \includegraphics{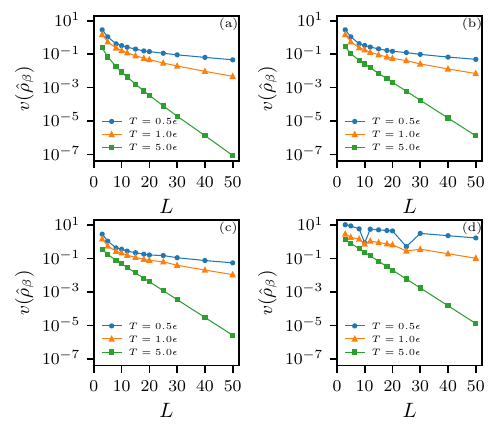}
    \caption{Variability $v(\hat{\rho}_{\beta})$ as a function of lead modes $L$ for different temperatures $T=0.5\epsilon, \epsilon, 5\epsilon$ (blue, orange, green) at different interaction strengths $U=0, 0.5\epsilon, \epsilon, 5\epsilon$ [panels (a), (b), (c), (d)]. We set $\epsilon=1, W=10$ and $\Gamma=1$.}
    \label{fig:variability_plot}
\end{figure}

Fig.~\ref{fig:variability_plot} displays the variability as a function of the number of lead modes $L$ for fixed system size $N=3$. It can be observed that, analogous to Fig.~\ref{fig:infidelity} for non-interacting systems, the variability monotonically decreases for increasing $L$. This is an indication that for increasing $L$, the extended system approaches a thermal fixed point, even in the presence of strong system-lead interactions.  We therefore conclude that a mesoscopic-lead description appropriately describes equilibrium states, even for more general systems that contain non-quadratic interactions and non-trivial system-bath correlations.

\subsection{Entropy production}
\label{sec:spohn}

\begin{figure}
    \centering
    \includegraphics{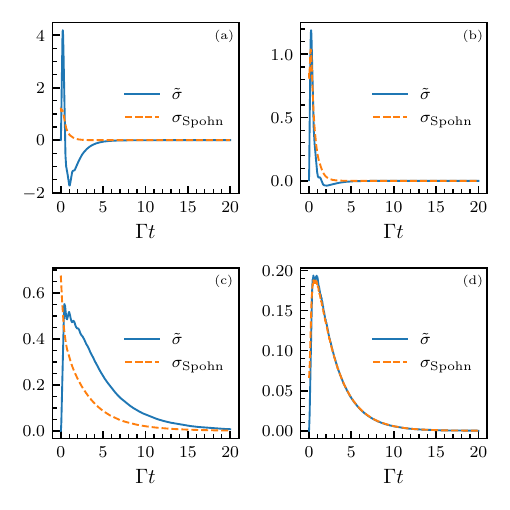}
    \caption{Comparison of the two different definitions of the entropy production rate $\tilde{\sigma}(t)$ and $\sigma_{\rm Spohn}(t)$ for the resonant level model for different temperatures and lead size. We set $\epsilon=1$, $\Gamma=1$ and $W=10$ in all plots. The temperatures and lead sizes are the following: (a) $T=0.1\epsilon$ and $L=5$; (b) $T=\epsilon$ and $L=5$; (c) $T=0.1 \epsilon$, and $L=100$; (d) $T=\epsilon$ and $L=100$. We see that in (d),  which corresponds to the high-temperature limit with a large number of modes in the lead, we get good agreement between the two definitions, as expected.}
    \label{fig:Spohn}
\end{figure}

Next we will seek to understand how the entropy production (defined over the entire extended system) behaves for a single bath in the transient regime before reaching the steady state.
To begin we must first define the currents flowing between the lead modes and their respective reservoirs (which we refer to as ``external currents''). The external energy and particle currents are defined, respectively, as
\begin{align}
    \begin{split}
    I^E(t) &= \Tr[\hat{H}\mathcal{L}\{\hat{\rho}_{SL}(t)\}] \quad\text{and}\\
    I^P(t) &= \Tr[\hat{N}\mathcal{L}\{\hat{\rho}_{SL}(t)\}].
    \end{split}
\end{align}
Furthermore we can define the external heat current as 
\begin{equation}
    I^Q(t) = I^E(t) - \mu I^P(t)\,.
\end{equation}
Following the definitions of the entropy production rate as Refs.~\cite{Esposito2009a, Reeb_2014, Landi_2021, Strasberg_tutorial_2021, Lacerda_2022}, as the difference between the dissipated heat current and the change in the instantaneous von Neumann entropy of the extended system, we can write down the expression for the entropy production rate as
\begin{equation}
\label{eq:external_entropy_production_rate}
    \tilde{\sigma}(t)= \dot{S}_{SL}(t) -  \beta I^{Q}(t)\,,
\end{equation}
where we are using the shorthand notation for the von Neumann entropy of the extended system $S(\hat{\rho}_{SL}(t)) = S_{SL}(t)$ and $\dot{S}_{SL}(t)$ is its time derivative.
Note this definition does not rely on the assumption that the fixed point of the extended system is thermal~\cite{Lacerda_2022}.

Since the extended system obeys Lindblad dynamics, we also consider Spohn's definition of the entropy production rate~\cite{Spohn_1978}
\begin{equation}\label{eq:spohn_entropy_production_rate}
    \sigma_\text{Spohn}(t) = -\frac{d}{d t}D(\hat{\rho}_{SL}(t) || \hat{\rho}_{ss}) \geq 0,
\end{equation}
which is non-negative for Markovian semigroup dynamics with a fixed point $\hat{\rho}_{ss}$: this follows from the contractive property of the relative entropy under CPTP maps~\cite{Spohn_1978}. If, furthermore, the fixed point is given by the thermal state~\eqref{eq:thermal}, it is easy to show that
\begin{equation}
    \sigma_{\rm Spohn}(t) = \tilde{\sigma}(t)\,.
\end{equation}

As discussed in the previous section, the fixed point of the dynamics is approximately thermal when the number of modes is large. Therefore, we expect that the entropy production rates computed from Eq.~\eqref{eq:external_entropy_production_rate} and Eq.~\eqref{eq:spohn_entropy_production_rate} will agree in the limit of a high number of lead modes, $L$, or high temperature. To showcase this, we consider the resonant-level model with a single site in the central system, i.e. $N=1$ for the Hamiltonian considered in Eq.~(\ref{eq:h_s}). In Fig.~\ref{fig:Spohn}, we compare these definitions of entropy production for different values of temperature and increasing number of modes, with the flat spectral density given by Eq.~(\ref{eq:wideband}). 
Agreement between $\sigma_\text{Spohn}(t)$ and $\tilde{\sigma}(t)$ improves as the number of lead modes $L$ is increased, with more rapid convergence at higher temperature. This is in accordance with expectations, since the accuracy of the mesoscopic-leads approach generally improves with increasing $L$, albeit with larger values of $L$ needed to achieve the same accuracy at smaller temperatures~\cite{Brenes2020,Lacerda_2022}. Recovery of the correct transient entropy production rate through the Spohn prescription, even for strong system-reservoir coupling, is a powerful feature of the mescosopic-leads approach.

Note that in Fig.~\ref{fig:Spohn} we have considered relatively small values of $L$ since the computation of the extended system entropy derivative $\dot{S}_{SL}(t)$ becomes badly conditioned as $L$ grows very large. This issue does not arise when focussing on entropic quantities for the system $S$ alone, which is the case for the internal entropy production that we discuss in the following section.

\subsection{Internal and external entropy production rates}
\label{sec:difference_between_internal_and_external}

\subsubsection{Entropic discrepancy with a single bath}

So far we have studied the thermalisation of the extended system $\hat{\rho}_{SL}(t)$ in the steady state and the transient entropy production rate for a single bath. However it is important to stress that, despite the extended system having Markovian dynamics, the central system itself may not.
Given that, in the limit of large number of lead modes, Eq.~\eqref{eq:external_entropy_production_rate} and Eq.~\eqref{eq:spohn_entropy_production_rate} provide equivalent results for the entropy production rate for the global dynamics, the next step is to understand how this rate is related to the entropy production rate of the dynamics of the reduced system $\sigma(t)$, that is of $\hat{\rho}_{S}(t) = {\rm Tr}_{L}[\rho_{SL}(t)]$.

As discussed in Ref.~\cite{Lacerda_2022}, the particle and energy from the bath into the system (which we loosely refer to as ``internal currents'') can be computed from the master equation as 
\begin{align}
    \begin{split}
    J^E &= i\langle [\hat{H}_{L}, \hat{H}_{{SL}}]\rangle + \Tr[\hat{H}_{{SL}}\mathcal{D}\{\hat{\rho}_{SL}\}]  \quad\text{and}\\
    J^P &= i\langle [\hat{N}_{L}, \hat{H}_{{SL}}]\rangle.
    \end{split}
\end{align}
Given these definitions, one can straightforwardly define the entropy production rate associated with internal system as
\begin{equation}
\label{eq:internal_entropy_production_rate}
    \sigma(t) = \dot{S}_{S}(t) - \beta J^Q(t),
\end{equation}
where $J^Q(t) = J^E(t) - \mu J^P(t)$, which is very similar in construction to the external entropy production rate Eq.~(\ref{eq:external_entropy_production_rate}).

We note that when the system is coupled to a single bath, both $\sigma(t)$ and $\tilde{\sigma}(t)$ must vanish in the steady state. However, they may differ in the transient regime. 
This is to be expected, as information is thrown away when tracing out the lead.
In fact, $\tilde{\sigma}(t) \approx \sigma_{\rm Spohn}(t)$ is positive according to Ineq.~\eqref{eq:spohn_entropy_production_rate}, while $\sigma(t)$ can be negative since the reduced dynamics of the system is non-Markovian.

To better understand the discrepancy between internal and external entropy production rates, we turn our attention to the cumulative entropy production, defined as
\begin{equation}
    \Sigma(t) = \int_0^t d t' \sigma(t') \quad \text{and} \quad \tilde{\Sigma}(t) = \int_0^t d t' \tilde{\sigma}(t').
\end{equation}
If we substitute in the expressions for the rates Eq.~(\ref{eq:external_entropy_production_rate}) and Eq.~(\ref{eq:internal_entropy_production_rate}), the difference between $\Sigma(t)$ and $\tilde{\Sigma}(t)$ is given by
\begin{align}
\label{entropy_difference}
    {\Sigma(t) - \tilde{\Sigma}(t)} =& \int_0^t d t'\left[(\dot{S}_{S} - \dot{S}_{SL})  \right. \nonumber \\
    &\left.-\beta \{(J^E - I^E) + \mu (J^P - I^P)\}\right]\,,
\end{align}
where we suppress time arguments for concision. 

We can now systematically evaluate each of these integrals; see Appendix~\ref{app:evaluating_the_difference} for details. Assuming that the system and lead are initially uncorrelated, the first term in Eq.~\eqref{entropy_difference} can be written as
\begin{align}
\label{eq:entropy_diff_init}
    \int_0^t d t'[\dot{S}_{S} - \dot{S}_{SL}] = I({S}: {L}) - \Delta S_{L} ,
\end{align}
where $\Delta S_{L} = S_{L}(t) - S_{L}(0)$ and the mutual information between system and lead is defined by
\begin{equation}
    \label{mutual_information}
    I(S:L) = S_{S}(t) + S_{L}(t) - S_{SL}(t).
\end{equation}
The second term in Eq.~\eqref{entropy_difference} is related to the change in the number of particles within the lead:
\begin{align}
\label{eq:particle_diff}
     \int_0^t d t'[I^P - J^P] = \Delta N_{L}\,,
\end{align}
where $\Delta N_{L} = {\rm Tr}[ \hat{N}_L\{\hat{\rho}_{SL}(t) - \hat{\rho}_{SL}(0)\}]$. The final term in Eq.~\eqref{entropy_difference} is related to the change in energy of the lead:
\begin{align}
\label{eq:energy_diff}
     \int_0^t d t'[I^E - J^E]=\Delta E_{L}\,,
\end{align}
where $\Delta E_{L} = {\rm Tr}[\hat{H}_L\{\hat{\rho}_{SL}(t) - \hat{\rho}_{SL}(0)\}]$. Bringing everything together, we obtain an expression for the difference in the entropy production of the central and extended systems,
\begin{equation}
\label{eq:entropy_production_difference}
    \Sigma(t) - \tilde{\Sigma}(t) = \beta \Delta F_L + I({L}: {S}),
\end{equation}
where $\Delta F_{L} = F_{L}(t) - F_{L}(0)$ is the change in the non-equilibrium free energy of the lead, which is defined as $F_{L}(t) = E_{L} - \mu N_{L} - TS_{L}$. From this equation, we can immediately derive two conditions which together suffice to ensure that $\Sigma(t) \approx \tilde{\Sigma}(t)$ in the steady-state: (i) the lead returns to its initial state, i.e., $\Delta F_L \approx 0$; (ii) the correlations between system and lead are negligible, i.e., $I({L}: {S}) \approx 0$. 

\begin{figure}
    \centering
    \includegraphics{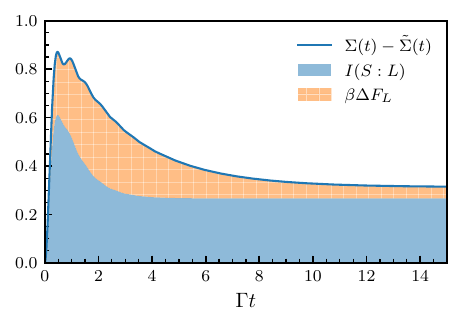}
    \caption{Difference in entropy production for the central and extended systems for a resonant-level model ($N=1$) coupled to a single bath. Shaded regions indicate the different contributions to the difference given by the RHS of Eq.~\eqref{eq:entropy_production_difference}. We again assume a flat spectrum with for the lead with the following parameters $\Gamma=\epsilon, T=\epsilon, \mu=0, L=100, \Delta t = 0.01, W=10\epsilon$.}
    \label{fig:entropy_difference_one_bath}
\end{figure}

In general, however, both terms on the right-hand side of Eq.~\eqref{eq:entropy_production_difference} are non-negative, and therefore $\Sigma(t) \geq \tilde{\Sigma}(t)$. In order to see this, note that the mutual information is non-negative, $I(S:L)\geq 0$. Moreover, assuming the initial state of the lead is thermal, its free energy cannot decrease. Indeed, one can write the free energy difference as a relative entropy \cite{Donald_1987}:
\begin{equation}
\label{eq:Mcdonald}
    \Delta F_L = \frac{1}{\beta}D\left(\hat{\rho}_{L }~\Big|\Big|~ \frac{e^{-\beta (\hat{H}_{L} - \mu \hat{N}_{L})}}{Z_{L}}\right) \geq 0,
\end{equation}
where we used the fact that $F_{L}(0) = T\log Z_{L}$. Therefore, we conclude that
\begin{equation}
\label{eq:entropy_diff}
    \Sigma(t) \geq \tilde{\Sigma}(t).
\end{equation}
This exact result holds independently of whether the system is interacting or non interacting. It expresses the fact that focussing on the central system (i.e.~tracing out the lead) represents a kind of coarse-graining, under which the entropy production can only increase~\cite{Esposito2012}.

We will now demonstrate this result by computing the entropy difference in Eq.~\eqref{eq:entropy_production_difference}, using the same resonant level model---we only have $N{=}1$ for the Hamiltonian considered in Eq.~(\ref{eq:h_s})---considered at the end of the previous Section.~\ref{sec:spohn}, which we depict in Fig.~\ref{fig:entropy_difference_one_bath}.
In this example we find perfect agreement between the LHS and RHS of Eq.~\eqref{eq:entropy_production_difference} where the mutual information accounts for the largest contribution to the difference. This can be attributed to  correlations created by the strong coupling between the system and the lead.  

\subsubsection{Entropic differences with multiple baths}

Consider a central system $S$ which is strongly coupled to $K$ distinct baths. In the mesoscopic-leads approach, each bath is described by a damped lead, and the joint state is given density operator $\hat{\rho}_{SL}$ and evolves according to Eq.~(\ref{eq:master_equation}).
In this setup we can define entropy production rate of the \emph{internal} system as 
\begin{equation}
    \sigma(t) = \dot{S}_{S}(t) - \sum_{\alpha=1}^{K}\beta_\alpha J^Q_\alpha(t),
\end{equation}
where $\dot{S}_{S}(t)$ is the time derivative of the von Neumann entropy of the reduced system, and $J^Q_\alpha = J^E_\alpha - \mu_\alpha J^P_\alpha$ is the internal heat current associated with the bath $\alpha$, with the internal energy and particle currents
\begin{align}
    \begin{split}
    J^E_\alpha(t) &= i\langle [\hat{H}_{L_\alpha}, \hat{H}_{{SL_\alpha}}]\rangle + \Tr[\hat{H}_{SL_\alpha}\mathcal{D}_\alpha\{\hat{\rho}_{SL}\}]  \quad\text{and}\\
    J^P_\alpha(t) &= i\langle [\hat{N}_{L_\alpha}, \hat{H}_{SL_\alpha}]\rangle.
    \end{split}
\end{align}
Here, $\hat{H}_{L_\alpha}$ is the Hamiltonian of lead $\alpha$, and $\hat{H}_{SL_\alpha}$ is the corresponding system-lead interaction.

The Spohn framework for entropy production generalises to multi-terminal setups in the following way~\cite{SpohnLebowitz1978}. We write the Lindblad equation for the extended system as
\begin{equation}
    \label{multi_bath_ME}
    \frac{d\hat{\rho}_{SL}}{dt} = -i[\hat{H},\hat{\rho}_{SL}] + \sum_{\alpha=1}^K \mathcal{D}_\alpha\{\hat{\rho}_{SL}\},
 \end{equation}
 where $\mathcal{D}_\alpha$ is the dissipation superoperator for lead $\alpha$. Let $\hat{\rho}_\alpha$ denote the steady state for that lead acting in isolation, i.e.
 \begin{equation}
     \label{rho_alpha}
     \mathcal{L}_\alpha \{\hat{\rho}_{\alpha}\} := -i[\hat{H},\hat{\rho}_\alpha] + \mathcal{D}_\alpha\{\hat{\rho}_\alpha\} = 0.
 \end{equation}
Therefore, contractivity of the relative entropy implies that~\cite{Spohn_1978}
\begin{equation}
    \Tr\left[\mathcal{L}_\alpha\{\hat{\rho}_{SL}(t)\}\left\{\ln \hat{\rho}_\alpha- \ln \hat{\rho}_{SL}(t) \right\}\right] \geq 0.
\end{equation}
Now, assuming that $\hat{\rho}_\alpha \propto e^{-\beta_\alpha(\hat{H}-\mu_\alpha\hat{N})}$ is a thermal state, one can show that
\begin{align}
    \label{external_sigma_multibath}
    \tilde{\sigma}(t) & := \dot{S}_{SL}(t) - \sum_{\alpha=1}^K \beta_\alpha I^Q_\alpha(t)\notag \\
    & =  \sum_{\alpha=1}^K \Tr\left[\mathcal{L}_\alpha\{\hat{\rho}_{SL}(t)\}\left\{\ln \hat{\rho}_\alpha- \ln \hat{\rho}_{SL}(t) \right\}\right] \geq 0.
\end{align}
The first equality defines the entropy production rate $\tilde{\sigma}(t)$ of the Markovian embedding, where $ I^Q_\alpha = I^E_\alpha - \mu_\alpha I^P_\alpha$ is the external heat current associated with bath $\alpha$, with the corresponding energy and particle currents
\begin{align}
\label{external_currents}
    \begin{split}
    I_\alpha^E(t) &= \Tr[\hat{H}\mathcal{D}_\alpha\{\hat{\rho}_{SL}(t)\}]  \quad\text{and}\\
    I_\alpha^P(t) &= \Tr[\hat{N}\mathcal{D}_\alpha\{\hat{\rho}_{SL}(t)\}].
    \end{split}
\end{align}

\begin{figure}
    \centering
    \includegraphics{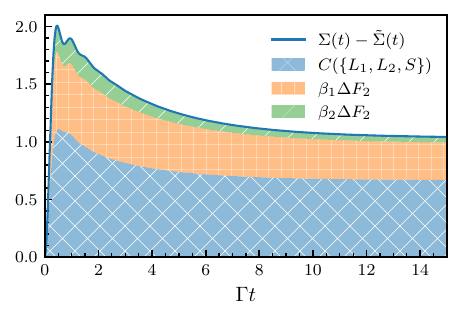}
    \caption{Entropy production difference between central and extended systems for a resonant-level model ($N=1$) coupled to two leads. Shaded regions denote the different contributions to the difference on the RHS of Eq.~\eqref{entropy_diff_multibath}. We again choose a flat spectrum for both leads with the following parameters $\Gamma=\epsilon, T_1=0.5\epsilon, T_2=\epsilon, \mu_1=\mu_2=0, L=100, \Delta t = 0.01, W=10\epsilon$. }
    \label{fig:entropy_diff_multiple}
\end{figure}

Repeating the analysis of the previous section, the difference between the entropy production of the extended and reduced systems can be shown to be
\begin{align}
\label{entropy_diff_multibath}
\Sigma(t) - \tilde{\Sigma}(t) &= \sum_{\alpha=1}^K \beta_\alpha \Delta F_{L_\alpha} + C(\{L_\alpha\},S),
\end{align}
where $\Delta F_{L_\alpha}$ is the change in non-equilibrium free energy of lead $\alpha$, and 
\begin{equation}
    C(\{L_\alpha\},S) = S_S(t) + \sum_{\alpha=1}^K S_{L_\alpha}(t) - S_{SL}(t)
\end{equation}
denotes the total correlations among the system and all the leads. As before, Eq.~\eqref{entropy_diff_multibath} holds assuming that the system and leads are mutually uncorrelated at $t=0$. As long as the leads begin in equilibrium, all terms on the RHS of Eq.~\eqref{entropy_diff_multibath} are non-negative, so we again conclude that $\Sigma(t) \geq \tilde{\Sigma}(t)$.

To illustrate this, we again use the resonant level model as an example, but now consider when it is connected to two terminals at different temperatures.
We can compute the entropic difference, as well as each individual term in Eq.~\eqref{entropy_diff_multibath} as a function of time, the results of which are depicted in Fig.~\ref{fig:entropy_diff_multiple} showing perfect agreement between the two. 
Analogously to Fig.~\ref{fig:entropy_difference_one_bath}, the largest contribution to the difference $\Sigma(t) - \Tilde{\Sigma}(t)$ is the total correlations between the leads and the system. Crucially, the difference between $\Sigma(t)$ and $\tilde{\Sigma}(t)$ tends to a constant in the non-equilibrium steady state, whereas the total entropy production scales with time. Therefore, either approach is appropriate for predicting thermodynamic quantities in the NESS.

\section{Discussion}

In this manuscript we have presented a comprehensive analysis of entropy production in the mesoscopic-leads approach to modelling open quantum systems. Firstly, we showed that, for a system strongly coupled to a single damped lead, the fixed point of the system approaches a thermal distribution for a sufficiently large number of lead modes, and also in the high temperature limit. 
In the non-interacting case, this was shown by computing the fidelity between the fixed point and a thermal ensemble, while in the interacting case we computed the variability of the thermal state using MPS methods. 

Using this result, we then showed that Spohn's framework for the entropy production of Markovian dynamical semigroups~\cite{Spohn_1978,SpohnLebowitz1978} can be used to define a non-negative entropy production rate at the level of the Markovian embedding. This was shown to agree with the standard definition of entropy production in terms of external currents, under the same conditions needed to obtain a thermal fixed point for a single lead, i.e.~many lead modes or high temperature. 

Finally, we examined the difference between the entropy production defined at the level of the central and extended systems, which are expressed in terms of internal and external currents, respectively. We showed that the former always exceeds the latter, presenting an elegant formula for the difference between the two. This result is significant, because it demonstrates that the internal currents computed within the mesoscopic-leads approach yield thermodynamically consistent predictions, even for a system strongly coupled to an arbitrary number of baths. While our results were illustrated using numerical examples in non-interacting systems, the assumptions underpinning these conclusions are valid for interacting and non-interacting systems alike.

Future work will extend these results to the realm of stochastic thermodynamics and determine whether the mesoscopic leads approach satisfies the quantum fluctuation theorems \cite{Manzano_2018, Manzano_2022} using quantum trajectories \cite{Bettmann2024}. Another interesting direction is to explore our results in the context of the periodically refreshed baths approach that has recently been introduced~\cite{archak1,archak2,archak3}. 

\begin{acknowledgments}
We are grateful to Laetitia Bettmann, Gabriel Landi, and Michael Zwolak for useful discussions.  This work was supported by the European Research Council Starting Grant ODYSSEY (Grant Agreement No.~758403) and the EPSRC-SFI joint project QuamNESS. J.G. is supported by a SFI-Royal Society University Research Fellowship. M.J.K. acknowledges the financial support from a Marie Sk\l odwoska-Curie Fellowship (Grant Agreement No.~101065974). The work of M.B.~has been supported by the Centre for Quantum Information and Quantum Control (CQIQC) at the University of Toronto and Universidad de Costa Rica (project Pry01-1750-2024). S.R.C. gratefully acknowledge financial support from UK's Engineering and Physical Sciences Research Council (EPSRC) under grant EP/T028424/1. M.T.M. is supported by a Royal Society-Science Foundation Ireland University Research Fellowship (URF\textbackslash R1\textbackslash 221571). This project is co-funded by the European Union (Quantum Flagship project ASPECTS, Grant Agreement No.~101080167). Views and opinions expressed are however those of the authors only and do not necessarily reflect those of the European Union, REA or UKRI. Neither
the European Union nor UKRI can be held responsible for them. 
\end{acknowledgments}

\bibliography{references}

\begin{thebibliography}{83}%
\makeatletter
\providecommand \@ifxundefined [1]{%
 \@ifx{#1\undefined}
}%
\providecommand \@ifnum [1]{%
 \ifnum #1\expandafter \@firstoftwo
 \else \expandafter \@secondoftwo
 \fi
}%
\providecommand \@ifx [1]{%
 \ifx #1\expandafter \@firstoftwo
 \else \expandafter \@secondoftwo
 \fi
}%
\providecommand \natexlab [1]{#1}%
\providecommand \enquote  [1]{``#1''}%
\providecommand \bibnamefont  [1]{#1}%
\providecommand \bibfnamefont [1]{#1}%
\providecommand \citenamefont [1]{#1}%
\providecommand \href@noop [0]{\@secondoftwo}%
\providecommand \href [0]{\begingroup \@sanitize@url \@href}%
\providecommand \@href[1]{\@@startlink{#1}\@@href}%
\providecommand \@@href[1]{\endgroup#1\@@endlink}%
\providecommand \@sanitize@url [0]{\catcode `\\12\catcode `\$12\catcode `\&12\catcode `\#12\catcode `\^12\catcode `\_12\catcode `\%12\relax}%
\providecommand \@@startlink[1]{}%
\providecommand \@@endlink[0]{}%
\providecommand \url  [0]{\begingroup\@sanitize@url \@url }%
\providecommand \@url [1]{\endgroup\@href {#1}{\urlprefix }}%
\providecommand \urlprefix  [0]{URL }%
\providecommand \Eprint [0]{\href }%
\providecommand \doibase [0]{http://dx.doi.org/}%
\providecommand \selectlanguage [0]{\@gobble}%
\providecommand \bibinfo  [0]{\@secondoftwo}%
\providecommand \bibfield  [0]{\@secondoftwo}%
\providecommand \translation [1]{[#1]}%
\providecommand \BibitemOpen [0]{}%
\providecommand \bibitemStop [0]{}%
\providecommand \bibitemNoStop [0]{.\EOS\space}%
\providecommand \EOS [0]{\spacefactor3000\relax}%
\providecommand \BibitemShut  [1]{\csname bibitem#1\endcsname}%
\let\auto@bib@innerbib\@empty
\bibitem [{\citenamefont {Landi}\ and\ \citenamefont {Paternostro}(2021)}]{Landi_2021}%
  \BibitemOpen
  \bibfield  {author} {\bibinfo {author} {\bibfnamefont {G.~T.}\ \bibnamefont {Landi}}\ and\ \bibinfo {author} {\bibfnamefont {M.}~\bibnamefont {Paternostro}},\ }\href {\doibase 10.1103/RevModPhys.93.035008} {\bibfield  {journal} {\bibinfo  {journal} {Rev. Mod. Phys.}\ }\textbf {\bibinfo {volume} {93}},\ \bibinfo {pages} {035008} (\bibinfo {year} {2021})}\BibitemShut {NoStop}%
\bibitem [{\citenamefont {Strasberg}\ and\ \citenamefont {Winter}(2021)}]{Strasberg_tutorial_2021}%
  \BibitemOpen
  \bibfield  {author} {\bibinfo {author} {\bibfnamefont {P.}~\bibnamefont {Strasberg}}\ and\ \bibinfo {author} {\bibfnamefont {A.}~\bibnamefont {Winter}},\ }\href {\doibase 10.1103/PRXQuantum.2.030202} {\bibfield  {journal} {\bibinfo  {journal} {PRX Quantum}\ }\textbf {\bibinfo {volume} {2}},\ \bibinfo {pages} {030202} (\bibinfo {year} {2021})}\BibitemShut {NoStop}%
\bibitem [{\citenamefont {Lebowitz}(1999)}]{Lebowitz_statistical_1999}%
  \BibitemOpen
  \bibfield  {author} {\bibinfo {author} {\bibfnamefont {J.~L.}\ \bibnamefont {Lebowitz}},\ }\href {\doibase 10.1103/RevModPhys.71.S346} {\bibfield  {journal} {\bibinfo  {journal} {Rev. Mod. Phys.}\ }\textbf {\bibinfo {volume} {71}},\ \bibinfo {pages} {S346} (\bibinfo {year} {1999})}\BibitemShut {NoStop}%
\bibitem [{\citenamefont {Talkner}\ and\ \citenamefont {H\"anggi}(2020)}]{Talkner2020}%
  \BibitemOpen
  \bibfield  {author} {\bibinfo {author} {\bibfnamefont {P.}~\bibnamefont {Talkner}}\ and\ \bibinfo {author} {\bibfnamefont {P.}~\bibnamefont {H\"anggi}},\ }\href {\doibase 10.1103/RevModPhys.92.041002} {\bibfield  {journal} {\bibinfo  {journal} {Rev. Mod. Phys.}\ }\textbf {\bibinfo {volume} {92}},\ \bibinfo {pages} {041002} (\bibinfo {year} {2020})}\BibitemShut {NoStop}%
\bibitem [{\citenamefont {Esposito}\ \emph {et~al.}(2010{\natexlab{a}})\citenamefont {Esposito}, \citenamefont {Lindenberg},\ and\ \citenamefont {den Broeck}}]{Esposito_2010}%
  \BibitemOpen
  \bibfield  {author} {\bibinfo {author} {\bibfnamefont {M.}~\bibnamefont {Esposito}}, \bibinfo {author} {\bibfnamefont {K.}~\bibnamefont {Lindenberg}}, \ and\ \bibinfo {author} {\bibfnamefont {C.~V.}\ \bibnamefont {den Broeck}},\ }\href {\doibase 10.1088/1367-2630/12/1/013013} {\bibfield  {journal} {\bibinfo  {journal} {New Journal of Physics}\ }\textbf {\bibinfo {volume} {12}},\ \bibinfo {pages} {013013} (\bibinfo {year} {2010}{\natexlab{a}})}\BibitemShut {NoStop}%
\bibitem [{\citenamefont {Reeb}\ and\ \citenamefont {Wolf}(2014)}]{Reeb_2014}%
  \BibitemOpen
  \bibfield  {author} {\bibinfo {author} {\bibfnamefont {D.}~\bibnamefont {Reeb}}\ and\ \bibinfo {author} {\bibfnamefont {M.~M.}\ \bibnamefont {Wolf}},\ }\href {\doibase 10.1088/1367-2630/16/10/103011} {\bibfield  {journal} {\bibinfo  {journal} {New Journal of Physics}\ }\textbf {\bibinfo {volume} {16}},\ \bibinfo {pages} {103011} (\bibinfo {year} {2014})}\BibitemShut {NoStop}%
\bibitem [{\citenamefont {Strasberg}\ \emph {et~al.}(2017)\citenamefont {Strasberg}, \citenamefont {Schaller}, \citenamefont {Brandes},\ and\ \citenamefont {Esposito}}]{Strasberg_quantum_2017}%
  \BibitemOpen
  \bibfield  {author} {\bibinfo {author} {\bibfnamefont {P.}~\bibnamefont {Strasberg}}, \bibinfo {author} {\bibfnamefont {G.}~\bibnamefont {Schaller}}, \bibinfo {author} {\bibfnamefont {T.}~\bibnamefont {Brandes}}, \ and\ \bibinfo {author} {\bibfnamefont {M.}~\bibnamefont {Esposito}},\ }\href {\doibase 10.1103/PhysRevX.7.021003} {\bibfield  {journal} {\bibinfo  {journal} {Phys. Rev. X}\ }\textbf {\bibinfo {volume} {7}},\ \bibinfo {pages} {021003} (\bibinfo {year} {2017})}\BibitemShut {NoStop}%
\bibitem [{\citenamefont {Ptaszy\ifmmode~\acute{n}\else \'{n}\fi{}ski}\ and\ \citenamefont {Esposito}(2019)}]{Ptaszynksi_entropy_2019}%
  \BibitemOpen
  \bibfield  {author} {\bibinfo {author} {\bibfnamefont {K.}~\bibnamefont {Ptaszy\ifmmode~\acute{n}\else \'{n}\fi{}ski}}\ and\ \bibinfo {author} {\bibfnamefont {M.}~\bibnamefont {Esposito}},\ }\href {\doibase 10.1103/PhysRevLett.123.200603} {\bibfield  {journal} {\bibinfo  {journal} {Phys. Rev. Lett.}\ }\textbf {\bibinfo {volume} {123}},\ \bibinfo {pages} {200603} (\bibinfo {year} {2019})}\BibitemShut {NoStop}%
\bibitem [{\citenamefont {Chen}\ \emph {et~al.}(2017)\citenamefont {Chen}, \citenamefont {Chen},\ and\ \citenamefont {Chen}}]{Chen_thermodynamic_2017}%
  \BibitemOpen
  \bibfield  {author} {\bibinfo {author} {\bibfnamefont {H.-B.}\ \bibnamefont {Chen}}, \bibinfo {author} {\bibfnamefont {G.-Y.}\ \bibnamefont {Chen}}, \ and\ \bibinfo {author} {\bibfnamefont {Y.-N.}\ \bibnamefont {Chen}},\ }\href {\doibase 10.1103/PhysRevA.96.062114} {\bibfield  {journal} {\bibinfo  {journal} {Phys. Rev. A}\ }\textbf {\bibinfo {volume} {96}},\ \bibinfo {pages} {062114} (\bibinfo {year} {2017})}\BibitemShut {NoStop}%
\bibitem [{\citenamefont {Li}(2017)}]{Wen_production_2017}%
  \BibitemOpen
  \bibfield  {author} {\bibinfo {author} {\bibfnamefont {S.-W.}\ \bibnamefont {Li}},\ }\href {\doibase 10.1103/PhysRevE.96.012139} {\bibfield  {journal} {\bibinfo  {journal} {Phys. Rev. E}\ }\textbf {\bibinfo {volume} {96}},\ \bibinfo {pages} {012139} (\bibinfo {year} {2017})}\BibitemShut {NoStop}%
\bibitem [{\citenamefont {Engelhardt}\ and\ \citenamefont {Schaller}(2018)}]{Engelhardt_2018}%
  \BibitemOpen
  \bibfield  {author} {\bibinfo {author} {\bibfnamefont {G.}~\bibnamefont {Engelhardt}}\ and\ \bibinfo {author} {\bibfnamefont {G.}~\bibnamefont {Schaller}},\ }\href {\doibase 10.1088/1367-2630/aaa38d} {\bibfield  {journal} {\bibinfo  {journal} {New Journal of Physics}\ }\textbf {\bibinfo {volume} {20}},\ \bibinfo {pages} {023011} (\bibinfo {year} {2018})}\BibitemShut {NoStop}%
\bibitem [{\citenamefont {Manzano}\ \emph {et~al.}(2018)\citenamefont {Manzano}, \citenamefont {Horowitz},\ and\ \citenamefont {Parrondo}}]{Manzano_2018}%
  \BibitemOpen
  \bibfield  {author} {\bibinfo {author} {\bibfnamefont {G.}~\bibnamefont {Manzano}}, \bibinfo {author} {\bibfnamefont {J.~M.}\ \bibnamefont {Horowitz}}, \ and\ \bibinfo {author} {\bibfnamefont {J.~M.}\ \bibnamefont {Parrondo}},\ }\href {\doibase 10.1103/physrevx.8.031037} {\bibfield  {journal} {\bibinfo  {journal} {Physical Review X}\ }\textbf {\bibinfo {volume} {8}} (\bibinfo {year} {2018}),\ 10.1103/physrevx.8.031037}\BibitemShut {NoStop}%
\bibitem [{\citenamefont {Santos}\ \emph {et~al.}(2019)\citenamefont {Santos}, \citenamefont {C{\'{e}}leri}, \citenamefont {Landi},\ and\ \citenamefont {Paternostro}}]{Santos_2019}%
  \BibitemOpen
  \bibfield  {author} {\bibinfo {author} {\bibfnamefont {J.~P.}\ \bibnamefont {Santos}}, \bibinfo {author} {\bibfnamefont {L.~C.}\ \bibnamefont {C{\'{e}}leri}}, \bibinfo {author} {\bibfnamefont {G.~T.}\ \bibnamefont {Landi}}, \ and\ \bibinfo {author} {\bibfnamefont {M.}~\bibnamefont {Paternostro}},\ }\href {\doibase 10.1038/s41534-019-0138-y} {\bibfield  {journal} {\bibinfo  {journal} {npj Quantum Information}\ }\textbf {\bibinfo {volume} {5}} (\bibinfo {year} {2019}),\ 10.1038/s41534-019-0138-y}\BibitemShut {NoStop}%
\bibitem [{\citenamefont {Bera}\ \emph {et~al.}(2019)\citenamefont {Bera}, \citenamefont {Riera}, \citenamefont {Lewenstein}, \citenamefont {Khanian},\ and\ \citenamefont {Winter}}]{Bera_2019}%
  \BibitemOpen
  \bibfield  {author} {\bibinfo {author} {\bibfnamefont {M.~N.}\ \bibnamefont {Bera}}, \bibinfo {author} {\bibfnamefont {A.}~\bibnamefont {Riera}}, \bibinfo {author} {\bibfnamefont {M.}~\bibnamefont {Lewenstein}}, \bibinfo {author} {\bibfnamefont {Z.~B.}\ \bibnamefont {Khanian}}, \ and\ \bibinfo {author} {\bibfnamefont {A.}~\bibnamefont {Winter}},\ }\href {\doibase 10.22331/q-2019-02-14-121} {\bibfield  {journal} {\bibinfo  {journal} {Quantum}\ }\textbf {\bibinfo {volume} {3}},\ \bibinfo {pages} {121} (\bibinfo {year} {2019})}\BibitemShut {NoStop}%
\bibitem [{\citenamefont {Woods}\ \emph {et~al.}(2014)\citenamefont {Woods}, \citenamefont {Groux}, \citenamefont {Chin}, \citenamefont {Huelga},\ and\ \citenamefont {Plenio}}]{Woods2014}%
  \BibitemOpen
  \bibfield  {author} {\bibinfo {author} {\bibfnamefont {M.~P.}\ \bibnamefont {Woods}}, \bibinfo {author} {\bibfnamefont {R.}~\bibnamefont {Groux}}, \bibinfo {author} {\bibfnamefont {A.~W.}\ \bibnamefont {Chin}}, \bibinfo {author} {\bibfnamefont {S.~F.}\ \bibnamefont {Huelga}}, \ and\ \bibinfo {author} {\bibfnamefont {M.~B.}\ \bibnamefont {Plenio}},\ }\href {\doibase 10.1063/1.4866769} {\bibfield  {journal} {\bibinfo  {journal} {J. Math. Phys.}\ }\textbf {\bibinfo {volume} {55}},\ \bibinfo {pages} {032101} (\bibinfo {year} {2014})}\BibitemShut {NoStop}%
\bibitem [{\citenamefont {Spohn}(1978)}]{Spohn_1978}%
  \BibitemOpen
  \bibfield  {author} {\bibinfo {author} {\bibfnamefont {H.}~\bibnamefont {Spohn}},\ }\href {\doibase 10.1063/1.523789} {\bibfield  {journal} {\bibinfo  {journal} {Journal of Mathematical Physics}\ }\textbf {\bibinfo {volume} {19}},\ \bibinfo {pages} {1227} (\bibinfo {year} {1978})}\BibitemShut {NoStop}%
\bibitem [{\citenamefont {Walls}(1970)}]{Walls1970}%
  \BibitemOpen
  \bibfield  {author} {\bibinfo {author} {\bibfnamefont {D.~F.}\ \bibnamefont {Walls}},\ }\href {\doibase 10.1007/bf01396784} {\bibfield  {journal} {\bibinfo  {journal} {Zeitschrift f\"{u}r Physik A Hadrons and nuclei}\ }\textbf {\bibinfo {volume} {234}},\ \bibinfo {pages} {231} (\bibinfo {year} {1970})}\BibitemShut {NoStop}%
\bibitem [{\citenamefont {Carmichael}\ and\ \citenamefont {Walls}(1973)}]{Carmichael1973}%
  \BibitemOpen
  \bibfield  {author} {\bibinfo {author} {\bibfnamefont {H.~J.}\ \bibnamefont {Carmichael}}\ and\ \bibinfo {author} {\bibfnamefont {D.~F.}\ \bibnamefont {Walls}},\ }\href {\doibase 10.1088/0305-4470/6/10/014} {\bibfield  {journal} {\bibinfo  {journal} {Journal of Physics A: Mathematical, Nuclear and General}\ }\textbf {\bibinfo {volume} {6}},\ \bibinfo {pages} {1552} (\bibinfo {year} {1973})}\BibitemShut {NoStop}%
\bibitem [{\citenamefont {Wichterich}\ \emph {et~al.}(2007)\citenamefont {Wichterich}, \citenamefont {Henrich}, \citenamefont {Breuer}, \citenamefont {Gemmer},\ and\ \citenamefont {Michel}}]{Wichterich_2007}%
  \BibitemOpen
  \bibfield  {author} {\bibinfo {author} {\bibfnamefont {H.}~\bibnamefont {Wichterich}}, \bibinfo {author} {\bibfnamefont {M.~J.}\ \bibnamefont {Henrich}}, \bibinfo {author} {\bibfnamefont {H.-P.}\ \bibnamefont {Breuer}}, \bibinfo {author} {\bibfnamefont {J.}~\bibnamefont {Gemmer}}, \ and\ \bibinfo {author} {\bibfnamefont {M.}~\bibnamefont {Michel}},\ }\href {\doibase 10.1103/PhysRevE.76.031115} {\bibfield  {journal} {\bibinfo  {journal} {Phys. Rev. E}\ }\textbf {\bibinfo {volume} {76}},\ \bibinfo {pages} {031115} (\bibinfo {year} {2007})}\BibitemShut {NoStop}%
\bibitem [{\citenamefont {Rivas}\ \emph {et~al.}(2010)\citenamefont {Rivas}, \citenamefont {Plato}, \citenamefont {Huelga},\ and\ \citenamefont {Plenio}}]{Rivas_2010}%
  \BibitemOpen
  \bibfield  {author} {\bibinfo {author} {\bibfnamefont {{\'{A}}.}~\bibnamefont {Rivas}}, \bibinfo {author} {\bibfnamefont {A.~D.~K.}\ \bibnamefont {Plato}}, \bibinfo {author} {\bibfnamefont {S.~F.}\ \bibnamefont {Huelga}}, \ and\ \bibinfo {author} {\bibfnamefont {M.~B.}\ \bibnamefont {Plenio}},\ }\href {\doibase 10.1088/1367-2630/12/11/113032} {\bibfield  {journal} {\bibinfo  {journal} {New Journal of Physics}\ }\textbf {\bibinfo {volume} {12}},\ \bibinfo {pages} {113032} (\bibinfo {year} {2010})}\BibitemShut {NoStop}%
\bibitem [{\citenamefont {Levy}\ and\ \citenamefont {Kosloff}(2014)}]{Levy2014}%
  \BibitemOpen
  \bibfield  {author} {\bibinfo {author} {\bibfnamefont {A.}~\bibnamefont {Levy}}\ and\ \bibinfo {author} {\bibfnamefont {R.}~\bibnamefont {Kosloff}},\ }\href {\doibase 10.1209/0295-5075/107/20004} {\bibfield  {journal} {\bibinfo  {journal} {Europhysics Letters}\ }\textbf {\bibinfo {volume} {107}},\ \bibinfo {pages} {20004} (\bibinfo {year} {2014})}\BibitemShut {NoStop}%
\bibitem [{\citenamefont {Barra}(2015)}]{Barra2015}%
  \BibitemOpen
  \bibfield  {author} {\bibinfo {author} {\bibfnamefont {F.}~\bibnamefont {Barra}},\ }\href {\doibase 10.1038/srep14873} {\bibfield  {journal} {\bibinfo  {journal} {Scientific Reports}\ }\textbf {\bibinfo {volume} {5}},\ \bibinfo {pages} {14873} (\bibinfo {year} {2015})}\BibitemShut {NoStop}%
\bibitem [{\citenamefont {Trushechkin}\ and\ \citenamefont {Volovich}(2016)}]{Trushechkin_2016}%
  \BibitemOpen
  \bibfield  {author} {\bibinfo {author} {\bibfnamefont {A.~S.}\ \bibnamefont {Trushechkin}}\ and\ \bibinfo {author} {\bibfnamefont {I.~V.}\ \bibnamefont {Volovich}},\ }\href {\doibase 10.1209/0295-5075/113/30005} {\bibfield  {journal} {\bibinfo  {journal} {{EPL} (Europhysics Letters)}\ }\textbf {\bibinfo {volume} {113}},\ \bibinfo {pages} {30005} (\bibinfo {year} {2016})}\BibitemShut {NoStop}%
\bibitem [{\citenamefont {González}\ \emph {et~al.}(2017)\citenamefont {González}, \citenamefont {Correa}, \citenamefont {Nocerino}, \citenamefont {Palao}, \citenamefont {Alonso},\ and\ \citenamefont {Adesso}}]{Gonzalez_2017}%
  \BibitemOpen
  \bibfield  {author} {\bibinfo {author} {\bibfnamefont {J.~O.}\ \bibnamefont {González}}, \bibinfo {author} {\bibfnamefont {L.~A.}\ \bibnamefont {Correa}}, \bibinfo {author} {\bibfnamefont {G.}~\bibnamefont {Nocerino}}, \bibinfo {author} {\bibfnamefont {J.~P.}\ \bibnamefont {Palao}}, \bibinfo {author} {\bibfnamefont {D.}~\bibnamefont {Alonso}}, \ and\ \bibinfo {author} {\bibfnamefont {G.}~\bibnamefont {Adesso}},\ }\href {\doibase 10.1142/S1230161217400108} {\bibfield  {journal} {\bibinfo  {journal} {Open Systems \& Information Dynamics}\ }\textbf {\bibinfo {volume} {24}},\ \bibinfo {pages} {1740010} (\bibinfo {year} {2017})}\BibitemShut {NoStop}%
\bibitem [{\citenamefont {Hofer}\ \emph {et~al.}(2017)\citenamefont {Hofer}, \citenamefont {Perarnau-Llobet}, \citenamefont {Miranda}, \citenamefont {Haack}, \citenamefont {Silva}, \citenamefont {Brask},\ and\ \citenamefont {Brunner}}]{Hofer_2017}%
  \BibitemOpen
  \bibfield  {author} {\bibinfo {author} {\bibfnamefont {P.~P.}\ \bibnamefont {Hofer}}, \bibinfo {author} {\bibfnamefont {M.}~\bibnamefont {Perarnau-Llobet}}, \bibinfo {author} {\bibfnamefont {L.~D.~M.}\ \bibnamefont {Miranda}}, \bibinfo {author} {\bibfnamefont {G.}~\bibnamefont {Haack}}, \bibinfo {author} {\bibfnamefont {R.}~\bibnamefont {Silva}}, \bibinfo {author} {\bibfnamefont {J.~B.}\ \bibnamefont {Brask}}, \ and\ \bibinfo {author} {\bibfnamefont {N.}~\bibnamefont {Brunner}},\ }\href {\doibase 10.1088/1367-2630/aa964f} {\bibfield  {journal} {\bibinfo  {journal} {New Journal of Physics}\ }\textbf {\bibinfo {volume} {19}},\ \bibinfo {pages} {123037} (\bibinfo {year} {2017})}\BibitemShut {NoStop}%
\bibitem [{\citenamefont {Naseem}\ \emph {et~al.}(2018)\citenamefont {Naseem}, \citenamefont {Xuereb},\ and\ \citenamefont {M\"ustecapl\ifmmode \imath \else \i \fi{}o\ifmmode~\breve{g}\else \u{g}\fi{}lu}}]{Naseem2018}%
  \BibitemOpen
  \bibfield  {author} {\bibinfo {author} {\bibfnamefont {M.~T.}\ \bibnamefont {Naseem}}, \bibinfo {author} {\bibfnamefont {A.}~\bibnamefont {Xuereb}}, \ and\ \bibinfo {author} {\bibfnamefont {O.~E.}\ \bibnamefont {M\"ustecapl\ifmmode \imath \else \i \fi{}o\ifmmode~\breve{g}\else \u{g}\fi{}lu}},\ }\href {\doibase 10.1103/PhysRevA.98.052123} {\bibfield  {journal} {\bibinfo  {journal} {Phys. Rev. A}\ }\textbf {\bibinfo {volume} {98}},\ \bibinfo {pages} {052123} (\bibinfo {year} {2018})}\BibitemShut {NoStop}%
\bibitem [{\citenamefont {Chiara}\ \emph {et~al.}(2018)\citenamefont {Chiara}, \citenamefont {Landi}, \citenamefont {Hewgill}, \citenamefont {Reid}, \citenamefont {Ferraro}, \citenamefont {Roncaglia},\ and\ \citenamefont {Antezza}}]{Chiara2018}%
  \BibitemOpen
  \bibfield  {author} {\bibinfo {author} {\bibfnamefont {G.~D.}\ \bibnamefont {Chiara}}, \bibinfo {author} {\bibfnamefont {G.}~\bibnamefont {Landi}}, \bibinfo {author} {\bibfnamefont {A.}~\bibnamefont {Hewgill}}, \bibinfo {author} {\bibfnamefont {B.}~\bibnamefont {Reid}}, \bibinfo {author} {\bibfnamefont {A.}~\bibnamefont {Ferraro}}, \bibinfo {author} {\bibfnamefont {A.~J.}\ \bibnamefont {Roncaglia}}, \ and\ \bibinfo {author} {\bibfnamefont {M.}~\bibnamefont {Antezza}},\ }\href {\doibase 10.1088/1367-2630/aaecee} {\bibfield  {journal} {\bibinfo  {journal} {New Journal of Physics}\ }\textbf {\bibinfo {volume} {20}},\ \bibinfo {pages} {113024} (\bibinfo {year} {2018})}\BibitemShut {NoStop}%
\bibitem [{\citenamefont {Mitchison}\ and\ \citenamefont {Plenio}(2018)}]{Mitchison_2018}%
  \BibitemOpen
  \bibfield  {author} {\bibinfo {author} {\bibfnamefont {M.~T.}\ \bibnamefont {Mitchison}}\ and\ \bibinfo {author} {\bibfnamefont {M.~B.}\ \bibnamefont {Plenio}},\ }\href {\doibase 10.1088/1367-2630/aa9f70} {\bibfield  {journal} {\bibinfo  {journal} {New Journal of Physics}\ }\textbf {\bibinfo {volume} {20}},\ \bibinfo {pages} {033005} (\bibinfo {year} {2018})}\BibitemShut {NoStop}%
\bibitem [{\citenamefont {Tupkary}\ \emph {et~al.}(2022)\citenamefont {Tupkary}, \citenamefont {Dhar}, \citenamefont {Kulkarni},\ and\ \citenamefont {Purkayastha}}]{Purkayastha2022}%
  \BibitemOpen
  \bibfield  {author} {\bibinfo {author} {\bibfnamefont {D.}~\bibnamefont {Tupkary}}, \bibinfo {author} {\bibfnamefont {A.}~\bibnamefont {Dhar}}, \bibinfo {author} {\bibfnamefont {M.}~\bibnamefont {Kulkarni}}, \ and\ \bibinfo {author} {\bibfnamefont {A.}~\bibnamefont {Purkayastha}},\ }\href {\doibase 10.1103/PhysRevA.105.032208} {\bibfield  {journal} {\bibinfo  {journal} {Phys. Rev. A}\ }\textbf {\bibinfo {volume} {105}},\ \bibinfo {pages} {032208} (\bibinfo {year} {2022})}\BibitemShut {NoStop}%
\bibitem [{\citenamefont {Imamoglu}(1994)}]{Imamoglu_1994}%
  \BibitemOpen
  \bibfield  {author} {\bibinfo {author} {\bibfnamefont {A.}~\bibnamefont {Imamoglu}},\ }\href {\doibase 10.1103/physreva.50.3650} {\bibfield  {journal} {\bibinfo  {journal} {Physical Review A}\ }\textbf {\bibinfo {volume} {50}},\ \bibinfo {pages} {3650} (\bibinfo {year} {1994})}\BibitemShut {NoStop}%
\bibitem [{\citenamefont {Garraway}(1997{\natexlab{a}})}]{Garraway_1997a}%
  \BibitemOpen
  \bibfield  {author} {\bibinfo {author} {\bibfnamefont {B.~M.}\ \bibnamefont {Garraway}},\ }\href {\doibase 10.1103/physreva.55.2290} {\bibfield  {journal} {\bibinfo  {journal} {Physical Review A}\ }\textbf {\bibinfo {volume} {55}},\ \bibinfo {pages} {2290} (\bibinfo {year} {1997}{\natexlab{a}})}\BibitemShut {NoStop}%
\bibitem [{\citenamefont {Garraway}(1997{\natexlab{b}})}]{Garraway_1997b}%
  \BibitemOpen
  \bibfield  {author} {\bibinfo {author} {\bibfnamefont {B.~M.}\ \bibnamefont {Garraway}},\ }\href {\doibase 10.1103/physreva.55.4636} {\bibfield  {journal} {\bibinfo  {journal} {Physical Review A}\ }\textbf {\bibinfo {volume} {55}},\ \bibinfo {pages} {4636} (\bibinfo {year} {1997}{\natexlab{b}})}\BibitemShut {NoStop}%
\bibitem [{\citenamefont {Sánchez}\ \emph {et~al.}(2006)\citenamefont {Sánchez}, \citenamefont {Stamenova}, \citenamefont {Sanvito}, \citenamefont {Bowler}, \citenamefont {Horsfield},\ and\ \citenamefont {Todorov}}]{Sanchez_2006}%
  \BibitemOpen
  \bibfield  {author} {\bibinfo {author} {\bibfnamefont {C.~G.}\ \bibnamefont {Sánchez}}, \bibinfo {author} {\bibfnamefont {M.}~\bibnamefont {Stamenova}}, \bibinfo {author} {\bibfnamefont {S.}~\bibnamefont {Sanvito}}, \bibinfo {author} {\bibfnamefont {D.~R.}\ \bibnamefont {Bowler}}, \bibinfo {author} {\bibfnamefont {A.~P.}\ \bibnamefont {Horsfield}}, \ and\ \bibinfo {author} {\bibfnamefont {T.~N.}\ \bibnamefont {Todorov}},\ }\href {\doibase 10.1063/1.2202329} {\bibfield  {journal} {\bibinfo  {journal} {The Journal of Chemical Physics}\ }\textbf {\bibinfo {volume} {124}},\ \bibinfo {pages} {214708} (\bibinfo {year} {2006})}\BibitemShut {NoStop}%
\bibitem [{\citenamefont {Subotnik}\ \emph {et~al.}(2009)\citenamefont {Subotnik}, \citenamefont {Hansen}, \citenamefont {Ratner},\ and\ \citenamefont {Nitzan}}]{Subotnik_2009}%
  \BibitemOpen
  \bibfield  {author} {\bibinfo {author} {\bibfnamefont {J.~E.}\ \bibnamefont {Subotnik}}, \bibinfo {author} {\bibfnamefont {T.}~\bibnamefont {Hansen}}, \bibinfo {author} {\bibfnamefont {M.~A.}\ \bibnamefont {Ratner}}, \ and\ \bibinfo {author} {\bibfnamefont {A.}~\bibnamefont {Nitzan}},\ }\href {\doibase 10.1063/1.3109898} {\bibfield  {journal} {\bibinfo  {journal} {The Journal of Chemical Physics}\ }\textbf {\bibinfo {volume} {130}},\ \bibinfo {pages} {144105} (\bibinfo {year} {2009})}\BibitemShut {NoStop}%
\bibitem [{\citenamefont {Dzhioev}\ and\ \citenamefont {Kosov}(2011)}]{Dzhioev_2011}%
  \BibitemOpen
  \bibfield  {author} {\bibinfo {author} {\bibfnamefont {A.~A.}\ \bibnamefont {Dzhioev}}\ and\ \bibinfo {author} {\bibfnamefont {D.~S.}\ \bibnamefont {Kosov}},\ }\href {\doibase 10.1063/1.3548065} {\bibfield  {journal} {\bibinfo  {journal} {The Journal of Chemical Physics}\ }\textbf {\bibinfo {volume} {134}},\ \bibinfo {pages} {044121} (\bibinfo {year} {2011})}\BibitemShut {NoStop}%
\bibitem [{\citenamefont {Ajisaka}\ \emph {et~al.}(2012)\citenamefont {Ajisaka}, \citenamefont {Barra}, \citenamefont {Mej{\'{\i}}a-Monasterio},\ and\ \citenamefont {Prosen}}]{Ajisaka_2012}%
  \BibitemOpen
  \bibfield  {author} {\bibinfo {author} {\bibfnamefont {S.}~\bibnamefont {Ajisaka}}, \bibinfo {author} {\bibfnamefont {F.}~\bibnamefont {Barra}}, \bibinfo {author} {\bibfnamefont {C.}~\bibnamefont {Mej{\'{\i}}a-Monasterio}}, \ and\ \bibinfo {author} {\bibfnamefont {T.}~\bibnamefont {Prosen}},\ }\href {\doibase 10.1103/physrevb.86.125111} {\bibfield  {journal} {\bibinfo  {journal} {Physical Review B}\ }\textbf {\bibinfo {volume} {86}} (\bibinfo {year} {2012}),\ 10.1103/physrevb.86.125111}\BibitemShut {NoStop}%
\bibitem [{\citenamefont {Ajisaka}\ and\ \citenamefont {Barra}(2013)}]{Ajisaka_2013}%
  \BibitemOpen
  \bibfield  {author} {\bibinfo {author} {\bibfnamefont {S.}~\bibnamefont {Ajisaka}}\ and\ \bibinfo {author} {\bibfnamefont {F.}~\bibnamefont {Barra}},\ }\href {\doibase 10.1103/physrevb.87.195114} {\bibfield  {journal} {\bibinfo  {journal} {Physical Review B}\ }\textbf {\bibinfo {volume} {87}} (\bibinfo {year} {2013}),\ 10.1103/physrevb.87.195114}\BibitemShut {NoStop}%
\bibitem [{\citenamefont {Arrigoni}\ \emph {et~al.}(2013)\citenamefont {Arrigoni}, \citenamefont {Knap},\ and\ \citenamefont {von~der Linden}}]{Arrigoni_2013}%
  \BibitemOpen
  \bibfield  {author} {\bibinfo {author} {\bibfnamefont {E.}~\bibnamefont {Arrigoni}}, \bibinfo {author} {\bibfnamefont {M.}~\bibnamefont {Knap}}, \ and\ \bibinfo {author} {\bibfnamefont {W.}~\bibnamefont {von~der Linden}},\ }\href {\doibase 10.1103/PhysRevLett.110.086403} {\bibfield  {journal} {\bibinfo  {journal} {Phys. Rev. Lett.}\ }\textbf {\bibinfo {volume} {110}},\ \bibinfo {pages} {086403} (\bibinfo {year} {2013})}\BibitemShut {NoStop}%
\bibitem [{\citenamefont {Dorda}\ \emph {et~al.}(2014)\citenamefont {Dorda}, \citenamefont {Nuss}, \citenamefont {von~der Linden},\ and\ \citenamefont {Arrigoni}}]{Dorda_2014}%
  \BibitemOpen
  \bibfield  {author} {\bibinfo {author} {\bibfnamefont {A.}~\bibnamefont {Dorda}}, \bibinfo {author} {\bibfnamefont {M.}~\bibnamefont {Nuss}}, \bibinfo {author} {\bibfnamefont {W.}~\bibnamefont {von~der Linden}}, \ and\ \bibinfo {author} {\bibfnamefont {E.}~\bibnamefont {Arrigoni}},\ }\href {\doibase 10.1103/PhysRevB.89.165105} {\bibfield  {journal} {\bibinfo  {journal} {Phys. Rev. B}\ }\textbf {\bibinfo {volume} {89}},\ \bibinfo {pages} {165105} (\bibinfo {year} {2014})}\BibitemShut {NoStop}%
\bibitem [{\citenamefont {Chen}\ \emph {et~al.}(2014)\citenamefont {Chen}, \citenamefont {Hansen},\ and\ \citenamefont {Franco}}]{Chen_2014}%
  \BibitemOpen
  \bibfield  {author} {\bibinfo {author} {\bibfnamefont {L.}~\bibnamefont {Chen}}, \bibinfo {author} {\bibfnamefont {T.}~\bibnamefont {Hansen}}, \ and\ \bibinfo {author} {\bibfnamefont {I.}~\bibnamefont {Franco}},\ }\href {\doibase 10.1021/jp505771f} {\bibfield  {journal} {\bibinfo  {journal} {The Journal of Physical Chemistry C}\ }\textbf {\bibinfo {volume} {118}},\ \bibinfo {pages} {20009} (\bibinfo {year} {2014})}\BibitemShut {NoStop}%
\bibitem [{\citenamefont {Zelovich}\ \emph {et~al.}(2014)\citenamefont {Zelovich}, \citenamefont {Kronik},\ and\ \citenamefont {Hod}}]{Zelovich_2014}%
  \BibitemOpen
  \bibfield  {author} {\bibinfo {author} {\bibfnamefont {T.}~\bibnamefont {Zelovich}}, \bibinfo {author} {\bibfnamefont {L.}~\bibnamefont {Kronik}}, \ and\ \bibinfo {author} {\bibfnamefont {O.}~\bibnamefont {Hod}},\ }\href {\doibase 10.1021/ct500135e} {\bibfield  {journal} {\bibinfo  {journal} {Journal of Chemical Theory and Computation}\ }\textbf {\bibinfo {volume} {10}},\ \bibinfo {pages} {2927} (\bibinfo {year} {2014})},\ \bibinfo {note} {pMID: 26588268}\BibitemShut {NoStop}%
\bibitem [{\citenamefont {Dorda}\ \emph {et~al.}(2015)\citenamefont {Dorda}, \citenamefont {Ganahl}, \citenamefont {Evertz}, \citenamefont {von~der Linden},\ and\ \citenamefont {Arrigoni}}]{Dorda_2015}%
  \BibitemOpen
  \bibfield  {author} {\bibinfo {author} {\bibfnamefont {A.}~\bibnamefont {Dorda}}, \bibinfo {author} {\bibfnamefont {M.}~\bibnamefont {Ganahl}}, \bibinfo {author} {\bibfnamefont {H.~G.}\ \bibnamefont {Evertz}}, \bibinfo {author} {\bibfnamefont {W.}~\bibnamefont {von~der Linden}}, \ and\ \bibinfo {author} {\bibfnamefont {E.}~\bibnamefont {Arrigoni}},\ }\href {\doibase 10.1103/PhysRevB.92.125145} {\bibfield  {journal} {\bibinfo  {journal} {Phys. Rev. B}\ }\textbf {\bibinfo {volume} {92}},\ \bibinfo {pages} {125145} (\bibinfo {year} {2015})}\BibitemShut {NoStop}%
\bibitem [{\citenamefont {Hod}\ \emph {et~al.}(2016)\citenamefont {Hod}, \citenamefont {Rodríguez-Rosario}, \citenamefont {Zelovich},\ and\ \citenamefont {Frauenheim}}]{Hod_2016}%
  \BibitemOpen
  \bibfield  {author} {\bibinfo {author} {\bibfnamefont {O.}~\bibnamefont {Hod}}, \bibinfo {author} {\bibfnamefont {C.~A.}\ \bibnamefont {Rodríguez-Rosario}}, \bibinfo {author} {\bibfnamefont {T.}~\bibnamefont {Zelovich}}, \ and\ \bibinfo {author} {\bibfnamefont {T.}~\bibnamefont {Frauenheim}},\ }\href {\doibase 10.1021/acs.jpca.5b12212} {\bibfield  {journal} {\bibinfo  {journal} {The Journal of Physical Chemistry A}\ }\textbf {\bibinfo {volume} {120}},\ \bibinfo {pages} {3278} (\bibinfo {year} {2016})},\ \bibinfo {note} {pMID: 26807992}\BibitemShut {NoStop}%
\bibitem [{\citenamefont {Gruss}\ \emph {et~al.}(2016)\citenamefont {Gruss}, \citenamefont {Velizhanin},\ and\ \citenamefont {Zwolak}}]{Gruss_2016}%
  \BibitemOpen
  \bibfield  {author} {\bibinfo {author} {\bibfnamefont {D.}~\bibnamefont {Gruss}}, \bibinfo {author} {\bibfnamefont {K.~A.}\ \bibnamefont {Velizhanin}}, \ and\ \bibinfo {author} {\bibfnamefont {M.}~\bibnamefont {Zwolak}},\ }\href {\doibase 10.1038/srep24514} {\bibfield  {journal} {\bibinfo  {journal} {Scientific Reports}\ }\textbf {\bibinfo {volume} {6}} (\bibinfo {year} {2016}),\ 10.1038/srep24514}\BibitemShut {NoStop}%
\bibitem [{\citenamefont {Schwarz}\ \emph {et~al.}(2016)\citenamefont {Schwarz}, \citenamefont {Goldstein}, \citenamefont {Dorda}, \citenamefont {Arrigoni}, \citenamefont {Weichselbaum},\ and\ \citenamefont {von Delft}}]{Schwarz_2016}%
  \BibitemOpen
  \bibfield  {author} {\bibinfo {author} {\bibfnamefont {F.}~\bibnamefont {Schwarz}}, \bibinfo {author} {\bibfnamefont {M.}~\bibnamefont {Goldstein}}, \bibinfo {author} {\bibfnamefont {A.}~\bibnamefont {Dorda}}, \bibinfo {author} {\bibfnamefont {E.}~\bibnamefont {Arrigoni}}, \bibinfo {author} {\bibfnamefont {A.}~\bibnamefont {Weichselbaum}}, \ and\ \bibinfo {author} {\bibfnamefont {J.}~\bibnamefont {von Delft}},\ }\href {\doibase 10.1103/PhysRevB.94.155142} {\bibfield  {journal} {\bibinfo  {journal} {Phys. Rev. B}\ }\textbf {\bibinfo {volume} {94}},\ \bibinfo {pages} {155142} (\bibinfo {year} {2016})}\BibitemShut {NoStop}%
\bibitem [{\citenamefont {Dorda}\ \emph {et~al.}(2017)\citenamefont {Dorda}, \citenamefont {Sorantin}, \citenamefont {von~der Linden},\ and\ \citenamefont {Arrigoni}}]{Dorda_2017}%
  \BibitemOpen
  \bibfield  {author} {\bibinfo {author} {\bibfnamefont {A.}~\bibnamefont {Dorda}}, \bibinfo {author} {\bibfnamefont {M.}~\bibnamefont {Sorantin}}, \bibinfo {author} {\bibfnamefont {W.}~\bibnamefont {von~der Linden}}, \ and\ \bibinfo {author} {\bibfnamefont {E.}~\bibnamefont {Arrigoni}},\ }\href {\doibase 10.1088/1367-2630/aa6ccc} {\bibfield  {journal} {\bibinfo  {journal} {New Journal of Physics}\ }\textbf {\bibinfo {volume} {19}},\ \bibinfo {pages} {063005} (\bibinfo {year} {2017})}\BibitemShut {NoStop}%
\bibitem [{\citenamefont {Elenewski}\ \emph {et~al.}(2017)\citenamefont {Elenewski}, \citenamefont {Gruss},\ and\ \citenamefont {Zwolak}}]{Elenewski_2017}%
  \BibitemOpen
  \bibfield  {author} {\bibinfo {author} {\bibfnamefont {J.~E.}\ \bibnamefont {Elenewski}}, \bibinfo {author} {\bibfnamefont {D.}~\bibnamefont {Gruss}}, \ and\ \bibinfo {author} {\bibfnamefont {M.}~\bibnamefont {Zwolak}},\ }\href {\doibase 10.1063/1.5000747} {\bibfield  {journal} {\bibinfo  {journal} {The Journal of Chemical Physics}\ }\textbf {\bibinfo {volume} {147}},\ \bibinfo {pages} {151101} (\bibinfo {year} {2017})}\BibitemShut {NoStop}%
\bibitem [{\citenamefont {Gruss}\ \emph {et~al.}(2017)\citenamefont {Gruss}, \citenamefont {Smolyanitsky},\ and\ \citenamefont {Zwolak}}]{Gruss_2017}%
  \BibitemOpen
  \bibfield  {author} {\bibinfo {author} {\bibfnamefont {D.}~\bibnamefont {Gruss}}, \bibinfo {author} {\bibfnamefont {A.}~\bibnamefont {Smolyanitsky}}, \ and\ \bibinfo {author} {\bibfnamefont {M.}~\bibnamefont {Zwolak}},\ }\href {\doibase 10.1063/1.4997022} {\bibfield  {journal} {\bibinfo  {journal} {The Journal of Chemical Physics}\ }\textbf {\bibinfo {volume} {147}},\ \bibinfo {pages} {141102} (\bibinfo {year} {2017})}\BibitemShut {NoStop}%
\bibitem [{\citenamefont {Zelovich}\ \emph {et~al.}(2017)\citenamefont {Zelovich}, \citenamefont {Hansen}, \citenamefont {Liu}, \citenamefont {Neaton}, \citenamefont {Kronik},\ and\ \citenamefont {Hod}}]{Zelovich_2017}%
  \BibitemOpen
  \bibfield  {author} {\bibinfo {author} {\bibfnamefont {T.}~\bibnamefont {Zelovich}}, \bibinfo {author} {\bibfnamefont {T.}~\bibnamefont {Hansen}}, \bibinfo {author} {\bibfnamefont {Z.-F.}\ \bibnamefont {Liu}}, \bibinfo {author} {\bibfnamefont {J.~B.}\ \bibnamefont {Neaton}}, \bibinfo {author} {\bibfnamefont {L.}~\bibnamefont {Kronik}}, \ and\ \bibinfo {author} {\bibfnamefont {O.}~\bibnamefont {Hod}},\ }\href {\doibase 10.1063/1.4976731} {\bibfield  {journal} {\bibinfo  {journal} {The Journal of Chemical Physics}\ }\textbf {\bibinfo {volume} {146}},\ \bibinfo {pages} {092331} (\bibinfo {year} {2017})}\BibitemShut {NoStop}%
\bibitem [{\citenamefont {Tamascelli}\ \emph {et~al.}(2018)\citenamefont {Tamascelli}, \citenamefont {Smirne}, \citenamefont {Huelga},\ and\ \citenamefont {Plenio}}]{Tamascelli_2018}%
  \BibitemOpen
  \bibfield  {author} {\bibinfo {author} {\bibfnamefont {D.}~\bibnamefont {Tamascelli}}, \bibinfo {author} {\bibfnamefont {A.}~\bibnamefont {Smirne}}, \bibinfo {author} {\bibfnamefont {S.}~\bibnamefont {Huelga}}, \ and\ \bibinfo {author} {\bibfnamefont {M.}~\bibnamefont {Plenio}},\ }\href {\doibase 10.1103/physrevlett.120.030402} {\bibfield  {journal} {\bibinfo  {journal} {Physical Review Letters}\ }\textbf {\bibinfo {volume} {120}} (\bibinfo {year} {2018}),\ 10.1103/physrevlett.120.030402}\BibitemShut {NoStop}%
\bibitem [{\citenamefont {Lemmer}\ \emph {et~al.}(2018)\citenamefont {Lemmer}, \citenamefont {Cormick}, \citenamefont {Tamascelli}, \citenamefont {Schaetz}, \citenamefont {Huelga},\ and\ \citenamefont {Plenio}}]{Lemmer_2018}%
  \BibitemOpen
  \bibfield  {author} {\bibinfo {author} {\bibfnamefont {A.}~\bibnamefont {Lemmer}}, \bibinfo {author} {\bibfnamefont {C.}~\bibnamefont {Cormick}}, \bibinfo {author} {\bibfnamefont {D.}~\bibnamefont {Tamascelli}}, \bibinfo {author} {\bibfnamefont {T.}~\bibnamefont {Schaetz}}, \bibinfo {author} {\bibfnamefont {S.~F.}\ \bibnamefont {Huelga}}, \ and\ \bibinfo {author} {\bibfnamefont {M.~B.}\ \bibnamefont {Plenio}},\ }\href {\doibase 10.1088/1367-2630/aac87d} {\bibfield  {journal} {\bibinfo  {journal} {New Journal of Physics}\ }\textbf {\bibinfo {volume} {20}},\ \bibinfo {pages} {073002} (\bibinfo {year} {2018})}\BibitemShut {NoStop}%
\bibitem [{\citenamefont {Oz}\ \emph {et~al.}(2019)\citenamefont {Oz}, \citenamefont {Hod},\ and\ \citenamefont {Nitzan}}]{Oz_2019}%
  \BibitemOpen
  \bibfield  {author} {\bibinfo {author} {\bibfnamefont {A.}~\bibnamefont {Oz}}, \bibinfo {author} {\bibfnamefont {O.}~\bibnamefont {Hod}}, \ and\ \bibinfo {author} {\bibfnamefont {A.}~\bibnamefont {Nitzan}},\ }\href {\doibase 10.1021/acs.jctc.9b00999} {\bibfield  {journal} {\bibinfo  {journal} {Journal of Chemical Theory and Computation}\ }\textbf {\bibinfo {volume} {16}},\ \bibinfo {pages} {1232} (\bibinfo {year} {2019})}\BibitemShut {NoStop}%
\bibitem [{\citenamefont {Chen}\ \emph {et~al.}(2019{\natexlab{a}})\citenamefont {Chen}, \citenamefont {Arrigoni},\ and\ \citenamefont {Galperin}}]{Chen_Galperin_2019}%
  \BibitemOpen
  \bibfield  {author} {\bibinfo {author} {\bibfnamefont {F.}~\bibnamefont {Chen}}, \bibinfo {author} {\bibfnamefont {E.}~\bibnamefont {Arrigoni}}, \ and\ \bibinfo {author} {\bibfnamefont {M.}~\bibnamefont {Galperin}},\ }\href {\doibase 10.1088/1367-2630/ab5ec5} {\bibfield  {journal} {\bibinfo  {journal} {New Journal of Physics}\ }\textbf {\bibinfo {volume} {21}},\ \bibinfo {pages} {123035} (\bibinfo {year} {2019}{\natexlab{a}})}\BibitemShut {NoStop}%
\bibitem [{\citenamefont {Zwolak}(2020)}]{Zwolak_2020}%
  \BibitemOpen
  \bibfield  {author} {\bibinfo {author} {\bibfnamefont {M.}~\bibnamefont {Zwolak}},\ }\href {\doibase 10.1063/5.0029223} {\bibfield  {journal} {\bibinfo  {journal} {The Journal of Chemical Physics}\ }\textbf {\bibinfo {volume} {153}},\ \bibinfo {pages} {224107} (\bibinfo {year} {2020})}\BibitemShut {NoStop}%
\bibitem [{\citenamefont {W{\'{o}}jtowicz}\ \emph {et~al.}(2020)\citenamefont {W{\'{o}}jtowicz}, \citenamefont {Elenewski}, \citenamefont {Rams},\ and\ \citenamefont {Zwolak}}]{Wojtowicz_2020}%
  \BibitemOpen
  \bibfield  {author} {\bibinfo {author} {\bibfnamefont {G.}~\bibnamefont {W{\'{o}}jtowicz}}, \bibinfo {author} {\bibfnamefont {J.~E.}\ \bibnamefont {Elenewski}}, \bibinfo {author} {\bibfnamefont {M.~M.}\ \bibnamefont {Rams}}, \ and\ \bibinfo {author} {\bibfnamefont {M.}~\bibnamefont {Zwolak}},\ }\href {\doibase 10.1103/physreva.101.050301} {\bibfield  {journal} {\bibinfo  {journal} {Physical Review A}\ }\textbf {\bibinfo {volume} {101}} (\bibinfo {year} {2020}),\ 10.1103/physreva.101.050301}\BibitemShut {NoStop}%
\bibitem [{\citenamefont {Chiang}\ and\ \citenamefont {Hsu}(2020)}]{Chiang_2020}%
  \BibitemOpen
  \bibfield  {author} {\bibinfo {author} {\bibfnamefont {T.-M.}\ \bibnamefont {Chiang}}\ and\ \bibinfo {author} {\bibfnamefont {L.-Y.}\ \bibnamefont {Hsu}},\ }\href {\doibase 10.1063/5.0007750} {\bibfield  {journal} {\bibinfo  {journal} {The Journal of Chemical Physics}\ }\textbf {\bibinfo {volume} {153}},\ \bibinfo {pages} {044103} (\bibinfo {year} {2020})}\BibitemShut {NoStop}%
\bibitem [{\citenamefont {Brenes}\ \emph {et~al.}(2020)\citenamefont {Brenes}, \citenamefont {Mendoza-Arenas}, \citenamefont {Purkayastha}, \citenamefont {Mitchison}, \citenamefont {Clark},\ and\ \citenamefont {Goold}}]{Brenes2020}%
  \BibitemOpen
  \bibfield  {author} {\bibinfo {author} {\bibfnamefont {M.}~\bibnamefont {Brenes}}, \bibinfo {author} {\bibfnamefont {J.~J.}\ \bibnamefont {Mendoza-Arenas}}, \bibinfo {author} {\bibfnamefont {A.}~\bibnamefont {Purkayastha}}, \bibinfo {author} {\bibfnamefont {M.~T.}\ \bibnamefont {Mitchison}}, \bibinfo {author} {\bibfnamefont {S.~R.}\ \bibnamefont {Clark}}, \ and\ \bibinfo {author} {\bibfnamefont {J.}~\bibnamefont {Goold}},\ }\href {\doibase 10.1103/PhysRevX.10.031040} {\bibfield  {journal} {\bibinfo  {journal} {Phys. Rev. X}\ }\textbf {\bibinfo {volume} {10}},\ \bibinfo {pages} {031040} (\bibinfo {year} {2020})}\BibitemShut {NoStop}%
\bibitem [{\citenamefont {Lotem}\ \emph {et~al.}(2020)\citenamefont {Lotem}, \citenamefont {Weichselbaum}, \citenamefont {von Delft},\ and\ \citenamefont {Goldstein}}]{Lotem_2020}%
  \BibitemOpen
  \bibfield  {author} {\bibinfo {author} {\bibfnamefont {M.}~\bibnamefont {Lotem}}, \bibinfo {author} {\bibfnamefont {A.}~\bibnamefont {Weichselbaum}}, \bibinfo {author} {\bibfnamefont {J.}~\bibnamefont {von Delft}}, \ and\ \bibinfo {author} {\bibfnamefont {M.}~\bibnamefont {Goldstein}},\ }\href {\doibase 10.1103/PhysRevResearch.2.043052} {\bibfield  {journal} {\bibinfo  {journal} {Phys. Rev. Research}\ }\textbf {\bibinfo {volume} {2}},\ \bibinfo {pages} {043052} (\bibinfo {year} {2020})}\BibitemShut {NoStop}%
\bibitem [{\citenamefont {Fugger}\ \emph {et~al.}(2020)\citenamefont {Fugger}, \citenamefont {Bauernfeind}, \citenamefont {Sorantin},\ and\ \citenamefont {Arrigoni}}]{Fugger_2020}%
  \BibitemOpen
  \bibfield  {author} {\bibinfo {author} {\bibfnamefont {D.~M.}\ \bibnamefont {Fugger}}, \bibinfo {author} {\bibfnamefont {D.}~\bibnamefont {Bauernfeind}}, \bibinfo {author} {\bibfnamefont {M.~E.}\ \bibnamefont {Sorantin}}, \ and\ \bibinfo {author} {\bibfnamefont {E.}~\bibnamefont {Arrigoni}},\ }\href {\doibase 10.1103/PhysRevB.101.165132} {\bibfield  {journal} {\bibinfo  {journal} {Phys. Rev. B}\ }\textbf {\bibinfo {volume} {101}},\ \bibinfo {pages} {165132} (\bibinfo {year} {2020})}\BibitemShut {NoStop}%
\bibitem [{\citenamefont {W\'ojtowicz}\ \emph {et~al.}(2021{\natexlab{a}})\citenamefont {W\'ojtowicz}, \citenamefont {Elenewski}, \citenamefont {Rams},\ and\ \citenamefont {Zwolak}}]{Wojtowicz_2021}%
  \BibitemOpen
  \bibfield  {author} {\bibinfo {author} {\bibfnamefont {G.}~\bibnamefont {W\'ojtowicz}}, \bibinfo {author} {\bibfnamefont {J.~E.}\ \bibnamefont {Elenewski}}, \bibinfo {author} {\bibfnamefont {M.~M.}\ \bibnamefont {Rams}}, \ and\ \bibinfo {author} {\bibfnamefont {M.}~\bibnamefont {Zwolak}},\ }\href {\doibase 10.1103/PhysRevB.104.165131} {\bibfield  {journal} {\bibinfo  {journal} {Phys. Rev. B}\ }\textbf {\bibinfo {volume} {104}},\ \bibinfo {pages} {165131} (\bibinfo {year} {2021}{\natexlab{a}})}\BibitemShut {NoStop}%
\bibitem [{\citenamefont {Elenewski}\ \emph {et~al.}(2021{\natexlab{a}})\citenamefont {Elenewski}, \citenamefont {Wójtowicz}, \citenamefont {Rams},\ and\ \citenamefont {Zwolak}}]{Elenewski_2021}%
  \BibitemOpen
  \bibfield  {author} {\bibinfo {author} {\bibfnamefont {J.~E.}\ \bibnamefont {Elenewski}}, \bibinfo {author} {\bibfnamefont {G.}~\bibnamefont {Wójtowicz}}, \bibinfo {author} {\bibfnamefont {M.~M.}\ \bibnamefont {Rams}}, \ and\ \bibinfo {author} {\bibfnamefont {M.}~\bibnamefont {Zwolak}},\ }\href {\doibase 10.1063/5.0065799} {\bibfield  {journal} {\bibinfo  {journal} {The Journal of Chemical Physics}\ }\textbf {\bibinfo {volume} {155}},\ \bibinfo {pages} {124117} (\bibinfo {year} {2021}{\natexlab{a}})}\BibitemShut {NoStop}%
\bibitem [{\citenamefont {Chen}\ \emph {et~al.}(2019{\natexlab{b}})\citenamefont {Chen}, \citenamefont {Cohen},\ and\ \citenamefont {Galperin}}]{Chen_Galperin_2019_2}%
  \BibitemOpen
  \bibfield  {author} {\bibinfo {author} {\bibfnamefont {F.}~\bibnamefont {Chen}}, \bibinfo {author} {\bibfnamefont {G.}~\bibnamefont {Cohen}}, \ and\ \bibinfo {author} {\bibfnamefont {M.}~\bibnamefont {Galperin}},\ }\href {\doibase 10.1103/PhysRevLett.122.186803} {\bibfield  {journal} {\bibinfo  {journal} {Phys. Rev. Lett.}\ }\textbf {\bibinfo {volume} {122}},\ \bibinfo {pages} {186803} (\bibinfo {year} {2019}{\natexlab{b}})}\BibitemShut {NoStop}%
\bibitem [{\citenamefont {Lacerda}\ \emph {et~al.}(2023)\citenamefont {Lacerda}, \citenamefont {Purkayastha}, \citenamefont {Kewming}, \citenamefont {Landi},\ and\ \citenamefont {Goold}}]{Lacerda_2022}%
  \BibitemOpen
  \bibfield  {author} {\bibinfo {author} {\bibfnamefont {A.~M.}\ \bibnamefont {Lacerda}}, \bibinfo {author} {\bibfnamefont {A.}~\bibnamefont {Purkayastha}}, \bibinfo {author} {\bibfnamefont {M.}~\bibnamefont {Kewming}}, \bibinfo {author} {\bibfnamefont {G.~T.}\ \bibnamefont {Landi}}, \ and\ \bibinfo {author} {\bibfnamefont {J.}~\bibnamefont {Goold}},\ }\href {\doibase 10.1103/PhysRevB.107.195117} {\bibfield  {journal} {\bibinfo  {journal} {Phys. Rev. B}\ }\textbf {\bibinfo {volume} {107}},\ \bibinfo {pages} {195117} (\bibinfo {year} {2023})}\BibitemShut {NoStop}%
\bibitem [{\citenamefont {Brenes}\ \emph {et~al.}(2023)\citenamefont {Brenes}, \citenamefont {Guarnieri}, \citenamefont {Purkayastha}, \citenamefont {Eisert}, \citenamefont {Segal},\ and\ \citenamefont {Landi}}]{Brenes_2022}%
  \BibitemOpen
  \bibfield  {author} {\bibinfo {author} {\bibfnamefont {M.}~\bibnamefont {Brenes}}, \bibinfo {author} {\bibfnamefont {G.}~\bibnamefont {Guarnieri}}, \bibinfo {author} {\bibfnamefont {A.}~\bibnamefont {Purkayastha}}, \bibinfo {author} {\bibfnamefont {J.}~\bibnamefont {Eisert}}, \bibinfo {author} {\bibfnamefont {D.}~\bibnamefont {Segal}}, \ and\ \bibinfo {author} {\bibfnamefont {G.}~\bibnamefont {Landi}},\ }\href {\doibase 10.1103/PhysRevB.108.L081119} {\bibfield  {journal} {\bibinfo  {journal} {Phys. Rev. B}\ }\textbf {\bibinfo {volume} {108}},\ \bibinfo {pages} {L081119} (\bibinfo {year} {2023})}\BibitemShut {NoStop}%
\bibitem [{\citenamefont {Reichental}\ \emph {et~al.}(2018)\citenamefont {Reichental}, \citenamefont {Klempner}, \citenamefont {Kafri},\ and\ \citenamefont {Podolsky}}]{Reichental2018}%
  \BibitemOpen
  \bibfield  {author} {\bibinfo {author} {\bibfnamefont {I.}~\bibnamefont {Reichental}}, \bibinfo {author} {\bibfnamefont {A.}~\bibnamefont {Klempner}}, \bibinfo {author} {\bibfnamefont {Y.}~\bibnamefont {Kafri}}, \ and\ \bibinfo {author} {\bibfnamefont {D.}~\bibnamefont {Podolsky}},\ }\href {\doibase 10.1103/PhysRevB.97.134301} {\bibfield  {journal} {\bibinfo  {journal} {Phys. Rev. B}\ }\textbf {\bibinfo {volume} {97}},\ \bibinfo {pages} {134301} (\bibinfo {year} {2018})}\BibitemShut {NoStop}%
\bibitem [{\citenamefont {Zanoci}\ \emph {et~al.}(2023)\citenamefont {Zanoci}, \citenamefont {Yoo},\ and\ \citenamefont {Swingle}}]{Zanoci2023}%
  \BibitemOpen
  \bibfield  {author} {\bibinfo {author} {\bibfnamefont {C.}~\bibnamefont {Zanoci}}, \bibinfo {author} {\bibfnamefont {Y.}~\bibnamefont {Yoo}}, \ and\ \bibinfo {author} {\bibfnamefont {B.}~\bibnamefont {Swingle}},\ }\href {\doibase 10.1103/PhysRevB.108.035156} {\bibfield  {journal} {\bibinfo  {journal} {Phys. Rev. B}\ }\textbf {\bibinfo {volume} {108}},\ \bibinfo {pages} {035156} (\bibinfo {year} {2023})}\BibitemShut {NoStop}%
\bibitem [{\citenamefont {Elenewski}\ \emph {et~al.}(2021{\natexlab{b}})\citenamefont {Elenewski}, \citenamefont {Wójtowicz}, \citenamefont {Rams},\ and\ \citenamefont {Zwolak}}]{Zwolak1:2021}%
  \BibitemOpen
  \bibfield  {author} {\bibinfo {author} {\bibfnamefont {J.~E.}\ \bibnamefont {Elenewski}}, \bibinfo {author} {\bibfnamefont {G.}~\bibnamefont {Wójtowicz}}, \bibinfo {author} {\bibfnamefont {M.~M.}\ \bibnamefont {Rams}}, \ and\ \bibinfo {author} {\bibfnamefont {M.}~\bibnamefont {Zwolak}},\ }\href {\doibase 10.1063/5.0065799} {\bibfield  {journal} {\bibinfo  {journal} {J. Chem. Phys.}\ }\textbf {\bibinfo {volume} {155}},\ \bibinfo {pages} {124117} (\bibinfo {year} {2021}{\natexlab{b}})}\BibitemShut {NoStop}%
\bibitem [{\citenamefont {W\'ojtowicz}\ \emph {et~al.}(2021{\natexlab{b}})\citenamefont {W\'ojtowicz}, \citenamefont {Elenewski}, \citenamefont {Rams},\ and\ \citenamefont {Zwolak}}]{Zwolak2:2021}%
  \BibitemOpen
  \bibfield  {author} {\bibinfo {author} {\bibfnamefont {G.}~\bibnamefont {W\'ojtowicz}}, \bibinfo {author} {\bibfnamefont {J.~E.}\ \bibnamefont {Elenewski}}, \bibinfo {author} {\bibfnamefont {M.~M.}\ \bibnamefont {Rams}}, \ and\ \bibinfo {author} {\bibfnamefont {M.}~\bibnamefont {Zwolak}},\ }\href {https://link.aps.org/doi/10.1103/PhysRevB.104.165131} {\bibfield  {journal} {\bibinfo  {journal} {Phys. Rev. B}\ }\textbf {\bibinfo {volume} {104}},\ \bibinfo {pages} {165131} (\bibinfo {year} {2021}{\natexlab{b}})}\BibitemShut {NoStop}%
\bibitem [{\citenamefont {Peschel}(2003)}]{Peschel_2003}%
  \BibitemOpen
  \bibfield  {author} {\bibinfo {author} {\bibfnamefont {I.}~\bibnamefont {Peschel}},\ }\href {\doibase 10.1088/0305-4470/36/14/101} {\bibfield  {journal} {\bibinfo  {journal} {Journal of Physics A: Mathematical and General}\ }\textbf {\bibinfo {volume} {36}},\ \bibinfo {pages} {L205} (\bibinfo {year} {2003})}\BibitemShut {NoStop}%
\bibitem [{\citenamefont {Bravyi}(2005)}]{Bravyi2005}%
  \BibitemOpen
  \bibfield  {author} {\bibinfo {author} {\bibfnamefont {S.}~\bibnamefont {Bravyi}},\ }\href@noop {} {\bibfield  {journal} {\bibinfo  {journal} {Quantum Info. Comput.}\ }\textbf {\bibinfo {volume} {5}},\ \bibinfo {pages} {216–238} (\bibinfo {year} {2005})}\BibitemShut {NoStop}%
\bibitem [{\citenamefont {Dhar}\ \emph {et~al.}(2012)\citenamefont {Dhar}, \citenamefont {Saito},\ and\ \citenamefont {H\"anggi}}]{Dhar2012}%
  \BibitemOpen
  \bibfield  {author} {\bibinfo {author} {\bibfnamefont {A.}~\bibnamefont {Dhar}}, \bibinfo {author} {\bibfnamefont {K.}~\bibnamefont {Saito}}, \ and\ \bibinfo {author} {\bibfnamefont {P.}~\bibnamefont {H\"anggi}},\ }\href {\doibase 10.1103/PhysRevE.85.011126} {\bibfield  {journal} {\bibinfo  {journal} {Phys. Rev. E}\ }\textbf {\bibinfo {volume} {85}},\ \bibinfo {pages} {011126} (\bibinfo {year} {2012})}\BibitemShut {NoStop}%
\bibitem [{\citenamefont {Banchi}\ \emph {et~al.}(2014)\citenamefont {Banchi}, \citenamefont {Giorda},\ and\ \citenamefont {Zanardi}}]{Banchi_quantum_2014}%
  \BibitemOpen
  \bibfield  {author} {\bibinfo {author} {\bibfnamefont {L.}~\bibnamefont {Banchi}}, \bibinfo {author} {\bibfnamefont {P.}~\bibnamefont {Giorda}}, \ and\ \bibinfo {author} {\bibfnamefont {P.}~\bibnamefont {Zanardi}},\ }\href {\doibase 10.1103/PhysRevE.89.022102} {\bibfield  {journal} {\bibinfo  {journal} {Phys. Rev. E}\ }\textbf {\bibinfo {volume} {89}},\ \bibinfo {pages} {022102} (\bibinfo {year} {2014})}\BibitemShut {NoStop}%
\bibitem [{\citenamefont {Bu{\v c}a}\ and\ \citenamefont {Prosen}(2012)}]{Buca2012}%
  \BibitemOpen
  \bibfield  {author} {\bibinfo {author} {\bibfnamefont {B.}~\bibnamefont {Bu{\v c}a}}\ and\ \bibinfo {author} {\bibfnamefont {T.}~\bibnamefont {Prosen}},\ }\href {\doibase 10.1088/1367-2630/14/7/073007} {\bibfield  {journal} {\bibinfo  {journal} {New Journal of Physics}\ }\textbf {\bibinfo {volume} {14}},\ \bibinfo {pages} {073007} (\bibinfo {year} {2012})}\BibitemShut {NoStop}%
\bibitem [{\citenamefont {Cirio}\ \emph {et~al.}(2023)\citenamefont {Cirio}, \citenamefont {Lambert}, \citenamefont {Liang}, \citenamefont {Kuo}, \citenamefont {Chen}, \citenamefont {Menczel}, \citenamefont {Funo},\ and\ \citenamefont {Nori}}]{Cirio:2023Pseudo}%
  \BibitemOpen
  \bibfield  {author} {\bibinfo {author} {\bibfnamefont {M.}~\bibnamefont {Cirio}}, \bibinfo {author} {\bibfnamefont {N.}~\bibnamefont {Lambert}}, \bibinfo {author} {\bibfnamefont {P.}~\bibnamefont {Liang}}, \bibinfo {author} {\bibfnamefont {P.-C.}\ \bibnamefont {Kuo}}, \bibinfo {author} {\bibfnamefont {Y.-N.}\ \bibnamefont {Chen}}, \bibinfo {author} {\bibfnamefont {P.}~\bibnamefont {Menczel}}, \bibinfo {author} {\bibfnamefont {K.}~\bibnamefont {Funo}}, \ and\ \bibinfo {author} {\bibfnamefont {F.}~\bibnamefont {Nori}},\ }\href {\doibase 10.1103/PhysRevResearch.5.033011} {\bibfield  {journal} {\bibinfo  {journal} {Phys. Rev. Res.}\ }\textbf {\bibinfo {volume} {5}},\ \bibinfo {pages} {033011} (\bibinfo {year} {2023})}\BibitemShut {NoStop}%
\bibitem [{\citenamefont {Esposito}\ \emph {et~al.}(2010{\natexlab{b}})\citenamefont {Esposito}, \citenamefont {Lindenberg},\ and\ \citenamefont {den Broeck}}]{Esposito2009a}%
  \BibitemOpen
  \bibfield  {author} {\bibinfo {author} {\bibfnamefont {M.}~\bibnamefont {Esposito}}, \bibinfo {author} {\bibfnamefont {K.}~\bibnamefont {Lindenberg}}, \ and\ \bibinfo {author} {\bibfnamefont {C.~V.}\ \bibnamefont {den Broeck}},\ }\href {\doibase 10.1088/1367-2630/12/1/013013} {\bibfield  {journal} {\bibinfo  {journal} {New Journal of Physics}\ }\textbf {\bibinfo {volume} {12}},\ \bibinfo {pages} {013013} (\bibinfo {year} {2010}{\natexlab{b}})}\BibitemShut {NoStop}%
\bibitem [{\citenamefont {Donald}(1987)}]{Donald_1987}%
  \BibitemOpen
  \bibfield  {author} {\bibinfo {author} {\bibfnamefont {M.~J.}\ \bibnamefont {Donald}},\ }\href {\doibase 10.1007/bf01009955} {\bibfield  {journal} {\bibinfo  {journal} {Journal of Statistical Physics}\ }\textbf {\bibinfo {volume} {49}},\ \bibinfo {pages} {81} (\bibinfo {year} {1987})}\BibitemShut {NoStop}%
\bibitem [{\citenamefont {Esposito}(2012)}]{Esposito2012}%
  \BibitemOpen
  \bibfield  {author} {\bibinfo {author} {\bibfnamefont {M.}~\bibnamefont {Esposito}},\ }\href {\doibase 10.1103/PhysRevE.85.041125} {\bibfield  {journal} {\bibinfo  {journal} {Phys. Rev. E}\ }\textbf {\bibinfo {volume} {85}},\ \bibinfo {pages} {041125} (\bibinfo {year} {2012})}\BibitemShut {NoStop}%
\bibitem [{\citenamefont {Spohn}\ and\ \citenamefont {Lebowitz}(1978)}]{SpohnLebowitz1978}%
  \BibitemOpen
  \bibfield  {author} {\bibinfo {author} {\bibfnamefont {H.}~\bibnamefont {Spohn}}\ and\ \bibinfo {author} {\bibfnamefont {J.~L.}\ \bibnamefont {Lebowitz}},\ }in\ \href {\doibase 10.1002/9780470142578.ch2} {\emph {\bibinfo {booktitle} {Advances in {{Chemical Physics}}}}},\ Vol.~\bibinfo {volume} {38},\ \bibinfo {editor} {edited by\ \bibinfo {editor} {\bibfnamefont {S.~A.}\ \bibnamefont {Rice}}}\ (\bibinfo  {publisher} {{Wiley}},\ \bibinfo {year} {1978})\ \bibinfo {edition} {1st}\ ed.,\ pp.\ \bibinfo {pages} {109--142}\BibitemShut {NoStop}%
\bibitem [{\citenamefont {Manzano}\ and\ \citenamefont {Zambrini}(2022)}]{Manzano_2022}%
  \BibitemOpen
  \bibfield  {author} {\bibinfo {author} {\bibfnamefont {G.}~\bibnamefont {Manzano}}\ and\ \bibinfo {author} {\bibfnamefont {R.}~\bibnamefont {Zambrini}},\ }\href {\doibase 10.1116/5.0079886} {\bibfield  {journal} {\bibinfo  {journal} {{AVS} Quantum Science}\ }\textbf {\bibinfo {volume} {4}},\ \bibinfo {pages} {025302} (\bibinfo {year} {2022})}\BibitemShut {NoStop}%
\bibitem [{\citenamefont {Bettmann}\ \emph {et~al.}(2024)\citenamefont {Bettmann}, \citenamefont {Kewming}, \citenamefont {Landi}, \citenamefont {Goold},\ and\ \citenamefont {Mitchison}}]{Bettmann2024}%
  \BibitemOpen
  \bibfield  {author} {\bibinfo {author} {\bibfnamefont {L.~P.}\ \bibnamefont {Bettmann}}, \bibinfo {author} {\bibfnamefont {M.~J.}\ \bibnamefont {Kewming}}, \bibinfo {author} {\bibfnamefont {G.~T.}\ \bibnamefont {Landi}}, \bibinfo {author} {\bibfnamefont {J.}~\bibnamefont {Goold}}, \ and\ \bibinfo {author} {\bibfnamefont {M.~T.}\ \bibnamefont {Mitchison}},\ }\href {\doibase 10.48550/arXiv.2404.06426} {\enquote {\bibinfo {title} {Quantum stochastic thermodynamics in the mesoscopic-leads formulation},}\ } (\bibinfo {year} {2024}),\ \Eprint {http://arxiv.org/abs/2404.06426} {arxiv:2404.06426 [cond-mat, physics:quant-ph]} \BibitemShut {NoStop}%
\bibitem [{\citenamefont {Purkayastha}\ \emph {et~al.}(2021)\citenamefont {Purkayastha}, \citenamefont {Guarnieri}, \citenamefont {Campbell}, \citenamefont {Prior},\ and\ \citenamefont {Goold}}]{archak1}%
  \BibitemOpen
  \bibfield  {author} {\bibinfo {author} {\bibfnamefont {A.}~\bibnamefont {Purkayastha}}, \bibinfo {author} {\bibfnamefont {G.}~\bibnamefont {Guarnieri}}, \bibinfo {author} {\bibfnamefont {S.}~\bibnamefont {Campbell}}, \bibinfo {author} {\bibfnamefont {J.}~\bibnamefont {Prior}}, \ and\ \bibinfo {author} {\bibfnamefont {J.}~\bibnamefont {Goold}},\ }\href {\doibase 10.1103/PhysRevB.104.045417} {\bibfield  {journal} {\bibinfo  {journal} {Phys. Rev. B}\ }\textbf {\bibinfo {volume} {104}},\ \bibinfo {pages} {045417} (\bibinfo {year} {2021})}\BibitemShut {NoStop}%
\bibitem [{\citenamefont {Purkayastha}\ \emph {et~al.}(2022)\citenamefont {Purkayastha}, \citenamefont {Guarnieri}, \citenamefont {Campbell}, \citenamefont {Prior},\ and\ \citenamefont {Goold}}]{archak2}%
  \BibitemOpen
  \bibfield  {author} {\bibinfo {author} {\bibfnamefont {A.}~\bibnamefont {Purkayastha}}, \bibinfo {author} {\bibfnamefont {G.}~\bibnamefont {Guarnieri}}, \bibinfo {author} {\bibfnamefont {S.}~\bibnamefont {Campbell}}, \bibinfo {author} {\bibfnamefont {J.}~\bibnamefont {Prior}}, \ and\ \bibinfo {author} {\bibfnamefont {J.}~\bibnamefont {Goold}},\ }\href {\doibase 10.22331/q-2022-09-08-801} {\bibfield  {journal} {\bibinfo  {journal} {{Quantum}}\ }\textbf {\bibinfo {volume} {6}},\ \bibinfo {pages} {801} (\bibinfo {year} {2022})}\BibitemShut {NoStop}%
\bibitem [{\citenamefont {W\'ojtowicz}\ \emph {et~al.}(2023)\citenamefont {W\'ojtowicz}, \citenamefont {Purkayastha}, \citenamefont {Zwolak},\ and\ \citenamefont {Rams}}]{archak3}%
  \BibitemOpen
  \bibfield  {author} {\bibinfo {author} {\bibfnamefont {G.}~\bibnamefont {W\'ojtowicz}}, \bibinfo {author} {\bibfnamefont {A.}~\bibnamefont {Purkayastha}}, \bibinfo {author} {\bibfnamefont {M.}~\bibnamefont {Zwolak}}, \ and\ \bibinfo {author} {\bibfnamefont {M.~M.}\ \bibnamefont {Rams}},\ }\href {\doibase 10.1103/PhysRevB.107.035150} {\bibfield  {journal} {\bibinfo  {journal} {Phys. Rev. B}\ }\textbf {\bibinfo {volume} {107}},\ \bibinfo {pages} {035150} (\bibinfo {year} {2023})}\BibitemShut {NoStop}%
\end{thebibliography}%
\bibliographystyle{apsrev4-1}

\appendix

\section{Computation of the relative entropy and fidelity from the covariance matrix}
\label{app:Fidelity and Entropy calculation}

Here we will show how one can efficiently compute the relative entropy in Eq.~(\ref{eq:Gaussian_relative_entropy}) and the Fidelity Eq.~(\ref{eq:fidelity}) for Gaussian states that can be uniquely defined by Eq.~(\ref{eq:Gaussian_rho}).
We will start with the definition of the von Neumann entropy 
\begin{equation}
    S(\hat{\rho}) = - {\rm Tr}(\hat{\rho} \log \hat{\rho})\,,
\end{equation}
where we can substitute Eq.~(\ref{eq:Gaussian_rho}) into the logarithm and then expand 
\begin{equation}
    S(\hat{\rho}) = \log Z + \sum_{ij}M_{ij}{\rm Tr}(\hat{\rho} \hat{d}_{i}^{\dagger}\hat{d}_{j})\,,
\end{equation}
which we can clearly see is just given by the covariance matrix Eq.~(\ref{eq:covariance})
yielding
\begin{equation}
    S(\hat{\rho}) = \log Z + {\rm Tr}(\mathbf{M} \mathbf{C})\,.
\end{equation}

The derivation of the relative entropy follows a similar logic. 
We begin with the definition 
\begin{equation}
    D(\hat{\rho}_{1}||\hat{\rho}_{2}) = - S(\hat{\rho}_{1}) - {\rm Tr}[\hat{\rho}_{1} \log \hat{\rho}_{2}]\,.
\end{equation}
If we expand out this expression using the same reasoning as above for the von Neumann entropy we simply obtain
\begin{equation}
    D(\hat{\rho}_{1}||\hat{\rho}_{2}) = \log \frac{Z_{2}}{Z_{1}} + {\rm Tr}\left([\mathbf{M}_{2} - \mathbf{M}_{1}]\mathbf{C}_{1}\right)\,,
\end{equation}
where $\mathbf{M}_j$, $\mathbf{C}_j$, and $Z_j$ pertain to the Gaussian state $\hat{\rho}_j$.

Lastly we will compute the Fidelity, which we will work through step by step. 
Recall that the Fidelity is defined by Eq.~(\ref{eq:fidelity}) which again reads
\begin{equation}
    F(\hat{\rho}_1, \hat{\rho}_2) = \Tr\left(\sqrt{\hat{\rho}_1 \hat{\rho}_2}\right)\,.
\end{equation}
The first step is to recognise
\begin{equation}
    \sqrt{\hat{\rho}_{1}\hat{\rho}_{2}} = \sqrt{\hat{\rho}}_{1}\sqrt{\hat{\rho}}_{2}\,,
\end{equation}
and can thus defined
\begin{equation}
    \sqrt{\hat{\rho}_{1}\hat{\rho}_{2}} = \frac{1}{\sqrt{Z_{1}Z_{2}}}e^{-\mathbf{d}^{\dagger}\mathbf{M}_{1}\mathbf{d}/2}e^{-\mathbf{d}^{\dagger}\mathbf{M}_{2}\mathbf{d}/2}\,,
\end{equation}
where $\mathbf{d}$ and $\mathbf{d}^{\dagger}$ are vectors containing the canonical fermionic operators $\hat{d}_i$ and $\hat{d}_i^{\dagger}$, respectively, for the $i$-th fermionic site.
From this expression we can now expand out the exponentials using the Baker-Campbell-Hausdorff formula resulting in 
\begin{align}
    e^{-\mathbf{d}^{\dagger}\mathbf{M}_{1}\mathbf{d}/2}e^{-\mathbf{d}^{\dagger}\mathbf{M}_{2}\mathbf{d}/2} = -\frac{\mathbf{d}^{\dagger}\mathbf{M}_{1}\mathbf{d}}{2} &-\frac{\mathbf{d}^{\dagger}\mathbf{M}_{2}\mathbf{d}}{2} \nonumber \\
    &- \frac{\mathbf{d}^{\dagger}[\mathbf{M}_{1}, M_{2}]\mathbf{d}}{4} + ...
\end{align}
where we have used the fact that terms $\mathbf{d}^{\dagger}\mathbf{M}_{1}\mathbf{d}^{\dagger} = \mathbf{d}\mathbf{M}_{1}\mathbf{d} = 0$ for Gaussian Fermionic systems.
With this expression, one can show that 
\begin{equation}
    e^{-\mathbf{d}^{\dagger}\mathbf{M}_{1}\mathbf{d}/2}e^{-\mathbf{d}^{\dagger}\mathbf{M}_{2}\mathbf{d}/2} = e^{-\mathbf{d}^{\dagger}\chi \mathbf{d}}\,,
\end{equation}
where $\chi = \log e^{-\mathbf{M_{1}}/2}e^{-\mathbf{M_{2}}/2}$\,.
Finally using the following relation
\begin{equation}
    {\rm Tr}\left(e^{-\mathbf{f}^{\dagger}\chi \mathbf{f}}\right) = {\rm det}\left(1 + e^{-\chi}\right)\,,
\end{equation}
therefore we can now explicitly write down the Fidelity as 
\begin{equation}
    F(\hat{\rho}_1, \hat{\rho}_2) = \frac{{\rm det}\left(1 + e^{-\mathbf{M_{1}}/2}e^{-\mathbf{M_{2}}/2}\right)}{\sqrt{Z_{1}Z_{2}}}\,.
\end{equation}

\section{Evaluating the difference integrals}
\label{app:evaluating_the_difference}
In this appendix, we work through the integrals described in Sec.~\ref{sec:difference_between_internal_and_external}.
We first evaluate the difference in the entropy production given by Eq.~(\ref{eq:entropy_diff})
\begin{align}
    \begin{split}
        \int_0^t d t'[\dot{S}_{S} - \dot{S}_{SL}] &= S_S(t) - S_S(0) - S_{SL}(t) + S_{SL}(0)\\
        &= S_S(t) - S_{SL}(t) + S_{L}(0),\\
    \end{split}
\end{align}
where we have used the fact that the system and lead are initially uncorrelated and thus $S_{SL}(0) = S_{L}(0) + S_{S}(0)$. Adding and subtracting $S_L(t)$, and making use of the mutual information $I(S:L) = S_{S}(t) + S_{L}(t) - S_{SL}(t)$, we obtain
\begin{align}
            \int_0^t d t'[\dot{S}_S - \dot{S}_{SL}] =I(S:L)- \Delta S_L.
    \end{align}
The second integral we want to evaluate is over the difference between the internal and external particle currents
\begin{align}
    \begin{split}
        \int_0^t d t'[I^P - J^P] &= \int_0^t d t' \Tr\left[\hat{N}_{L}\mathcal{D}\{\hat{\rho}\} -i[\hat{N}_{L}, \hat{H}_{SL}]\hat{\rho}\right] \\
        &= \int_0^t d t' \Tr\left[\hat{N}_{L}\mathcal{D}\{\hat{\rho}\} -i[\hat{N}_{L}, \hat{H}]\hat{\rho}\right]\\
        &= \int_0^t d t' \Tr\left[\hat{N}_{L}\left(\underbrace{\mathcal{D}\{\hat{\rho}\} - i[\hat{H}, \hat{\rho}]}_{d \hat{\rho}/d t'}\right)\right]\\
        &= \int_0^t d t' \frac{d}{d t'}\langle \hat{N}_{L}\rangle_{t'}\\
        &= \langle \hat{N}_{L}\rangle_t - \langle \hat{N}_{L}\rangle_0\\
        &= \Delta N_{L}
    \end{split}
\end{align}
The third integral is over the difference between the internal and external energy currents 
\begin{align}
    \begin{split}
        \int_0^t d t'[I^E - J^E] &=\int_0^t d t' \Tr\left[(\hat{H}_{L} + \hat{H}_{SL})\mathcal{D}\{\hat{\rho}\} \right. 
        \\
        &\qquad \quad\left.-i[\hat{H}_{L}, \hat{H}_{SL}]\hat{\rho}- \hat{H}_{SL}\mathcal{D}\{\hat{\rho}\}\right]\\
        &= \int_0^t d t' \Tr\left[\hat{H}_{L} \mathcal{D}\{\hat{\rho}\} -i[\hat{H}_{L}, \hat{H}_{SL}]\hat{\rho}\right]\\
        &= \int_0^t d t' \Tr\left[\hat{H}_{L}(\mathcal{D}\{\hat{\rho}\} -i[\hat{H}, \hat{\rho}])\right]\\
        &= \int_0^t d t' \Tr\left[\hat{H}_{L}\frac{d \hat{\rho}}{d t'}\right]\\
        &= \langle \hat{H}_{L}\rangle_t - \langle \hat{H}_{L}\rangle_0\\
        &= \Delta E_{L}.
    \end{split}
\end{align}

\end{document}